\documentclass[journal]{IEEEtran}
\usepackage[english]{babel} 
\usepackage{amsmath,amssymb,amsfonts}
\usepackage{graphicx}
\usepackage{caption}
\usepackage{booktabs}
\usepackage{xcolor}
\usepackage{physics}
\usepackage{tikz}
\usepackage{comment}
\usetikzlibrary{quantikz}
\usepackage{blindtext}
\usepackage[ruled,vlined]{algorithm2e}
\usepackage{xcolor}
\usepackage{soul}
\SetKw{KwBy}{by}
\SetKw{KwTo}{to}
\SetKwComment{Comment}{/* }{ */}
\usepackage{float}
\usepackage{balance}
\usepackage{ftnxtra}
\usepackage{fnpos}
\usepackage{caption}
\usepackage{subcaption}
\usetikzlibrary{shapes.geometric, arrows}

\tikzstyle{startstop} = [rectangle, rounded corners, minimum width=0.3\columnwidth, minimum height=1cm,text centered, draw=white, fill=white]

\tikzstyle{process} = [rectangle, minimum width=0.4\columnwidth, text width = 0.4\columnwidth, minimum height=1cm, text centered, draw=black, fill=orange!60]

\tikzstyle{process2} = [rectangle, minimum width=0.2\columnwidth, text width = 0.2\columnwidth, minimum height=1cm, text centered, draw=black, fill=orange!30]

\tikzstyle{arrow} = [thick,->,>=stealth]

\tikzstyle{textbox} = [rectangle, rounded corners, minimum width=0.1\columnwidth, minimum height=1cm,text centered, draw=white, fill=white]

\tikzstyle{inout} = [rectangle, rounded corners, minimum width=0.1\columnwidth, text width = 0.05\columnwidth, minimum height=1cm,text centered, draw=black, fill=orange!60]

\tikzstyle{gate} = [rectangle, rounded corners, minimum width=0.05\columnwidth, text width = 0.1\columnwidth, minimum height=1cm,text centered, draw=black, fill=white]

\tikzstyle{hidden} = [rectangle, rounded corners, minimum width=0.1\columnwidth, text width = 0.05\columnwidth, minimum height=1cm,text centered, draw=black, fill=red!60]

\tikzstyle{met} = [circle, minimum width=0.05\columnwidth, text width = 0.05\columnwidth, minimum height=1cm,text centered, draw=black, fill=white]

\begin{document}
\bstctlcite{IEEEexample:BSTcontrol}

\title{On Circuit-based Hybrid Quantum Neural Networks for Remote Sensing Imagery Classification}
\author{Alessandro~Sebastianelli,~\IEEEmembership{Student Member,~IEEE,} Daniela~A.~Zaidenberg,~\IEEEmembership{Student Member,~IEEE,} Dario~Spiller,~\IEEEmembership{Member,~IEEE,}~Bertrand~Le~Saux,~\IEEEmembership{Member,~IEEE,}~and~Silvia~L.~Ullo,~\IEEEmembership{Senior~Member,~IEEE}
\thanks{A. Sebastianelli and S. L. Ullo are with the Engineering Department, University of Sannio, Benevento, Italy, email: \{sebastianelli, ullo\}@unisannio.it}
\thanks{D. A. Zaidenberg is with the Massachusetts Institute of Technology, Boston, USA, email: dzaiden@mit.edu}
\thanks{D. Spiller and B. Le Saux are with the European Space Agency, $\Phi$-lab, Frascati, Italy, email: \{dario.spiller, bertrand.lesaux\}@esa.int}}
\maketitle
\begin{abstract}
This article aims to investigate how circuit-based hybrid Quantum Convolutional Neural Networks (QCNNs) can be successfully employed as image classifiers in the context of remote sensing. The hybrid QCNNs enrich the classical architecture of CNNs by introducing a quantum layer within a standard neural network. The novel QCNN proposed in this work is applied to the Land Use and Land Cover (LULC) classification, chosen as an Earth Observation (EO) use case, and tested on the EuroSAT dataset used as reference benchmark. The results of the multiclass classification prove the effectiveness of the presented approach, by demonstrating that the QCNN performances are higher than the classical counterparts. Moreover, investigation of various quantum circuits shows that the ones exploiting quantum entanglement achieve the best classification scores. This study underlines the potentialities of applying quantum computing to an EO case study and provides the theoretical and experimental background for futures investigations.
\end{abstract}

\section{Introduction}

\IEEEPARstart{E}{arth} Observation (EO) has consistently leveraged technological and computational advances helping in develop novel techniques to characterize and model the human environment~\cite{Rodriguez-Donaire2020,Sudmanns2019,Denis2017}.  Given that many remote sensing missions are currently operative, carrying on board multispectral, hyperspectral, and radar sensors, and the improved capabilities in transmitting and saving a continuously increasing number of images, nowadays estimated in over 150 terabytes per day~\cite{ESA_AI_EO}, the amount of data from EO applications has reached impressive volumes so that it is referred to as Big Data. At the same time, advances in computational technologies and analysis methodologies have also progressed to accommodate larger and higher-resolution datasets.
Image classification techniques are constantly being improved to keep up with the ever expanding stream of Big Data, and as a consequence Artificial Intelligence (AI) techniques are becoming increasingly necessary tools~\cite{Mathieu_book, del2021artificial}. 

Given the need to help expand the processing techniques to deal with this high resolution Big Data, EO is now looking towards new and innovative computation technologies~\cite{Riedel2021}. This is where Quantum Computing (QC) will play a fundamental role~\cite{NAP25196}.
Today, there  is a number of differing quantum devices, such as programmable superconducting processors~\cite{quantum_supremacy}, quantum annealers~\cite{McGeoch2020}, and photonic quantum computers~\cite{zhong2020quantum}. However, QC still presents some technological limitations, as reported in ~\cite{shettell2021practical} with a special concern with noise and limited error correction. Specific algorithms, namely the Noisy Intermediate-Scale Quantum Computing (NISQ) algorithms, have been designed to tackle these issues~\cite{bharti2021noisy}.

Quantum computers promise to efficiently solve important problems that are intractable on a conventional computer. For instance, in quantum systems, due to the exponentially growing physical dimensions, finding the eigenvalues of certain operators is one such intractable problem, which can be solved by combining a highly reconfigurable photonic quantum processor with a conventional computer~\cite{Peruzzo_2014,fedorov2021vqe}.


Another example is the case of the Variational Quantum Eigensolver (VQE) algorithm used to solve combinatorial optimization problems like finding the ground state energy of a molecule. The algorithm finds a bound to the lowest eigenenergy of a given Hamiltonian~\cite{fedorov2021vqe}. This is, in essence, a kind of cost function which is defined by the expectation of the molecular Hamiltonian of a given prepared eigenstate. The goal of the VQE is to minimize this cost function by varying the parameters $\theta$ used to prepare the ansatz eigenstate often representative of a molecule. This hybrid algorithm prepares and determines eigenenergies through quantum circuits, and then it varies the parameter classically. By iterating through these classical variations and quantum calculations, a hybrid minimization process is established~\cite{Peruzzo_2014}. This approximation of critical minima is analogous to the gradient descent. 

In QC  a qubit or quantum bit is the basic unit of quantum information, i.e. the quantum version of the classic binary bit. A qubit is one of the simplest quantum systems which displays the peculiarity of quantum mechanics. Indeed, it is a two-state quantum-mechanical system, e.g. an electron in two possible  levels (spin up and spin down), or a single photon in one of the two possible states (vertical and horizontal polarization). 
While in a classical system a bit can be in one state or the other, qubit exists in a coherent superposition of both states simultaneously, a property that is fundamental to quantum mechanics. Quantum computers utilize  the principles of superposition and entanglement to streamline computation~\cite{rieffel2011quantum,Kaye_intro_quantum,nielsen2011}. For every $n$ qubits, $2^n$ possible states can be represented. This is an exponential improvement with respect to the classical systems which can only represent n states for every n bits. Moreover, quantum systems exist in a high dimensional space, known as a Hilbert space, whose inherent properties lend themselves to a complex linear optimization. 

The application of quantum technology for remote sensing has been considered for at least the last 20 years.
In~\cite{Bi2019}, an active imaging information transmission technology for satellite-borne quantum remote sensing is proposed,  providing solutions and technical basis for realizing active imaging technology relying on quantum mechanics principles. 
Another application discussed in literature is related to  inteferometric synthetic aperture radars~\cite{Otgonbaatar2021a, Otgonbaatar2021b}. In the first work
Otgonbaatar and Datcu describe a residue connection problem in the phase unwrapping procedure as quadratic unconstrained binary optimization problem which is  solved by using the D-Wave quantum annealer. The same authors in~\cite{Otgonbaatar2021b} present a quantum annealer application for subset feature selection and the classification of hyperspectral images.

The research presented in this article focuses on the possibility to use quantum computers to enhance the performances of Machine Learning (ML) algorithms when applied to Land Use and Land Cover (LULC) classification, chosen as an EO use case. The results of the multiclass novel QCNN  classifier  prove  the  effectiveness  of  the  proposed approach, able to achieve better results that standard models of comparable complexity and on-par results with best standard models of the state of the art.

It is worth to highlight that only very few works have addressed the  application of Quantum Machine Learning (QML) to remote sensing in the current state of the art.  For instance, quantum computers and convolutional neural networks (CNNs) are considered together for accelerating geospatial data processing in~\cite{henderson2020methods}, where quanvolutional layers~\cite{henderson2019quanvolutional} are used. These layers contain several quanvolutional filters that transform the input data into different output feature maps by using a number of random quantum circuits, in an analogous way to standard convolutional networks. Quantum circuit-based neural network classifiers for multi-spectral land cover classification have been introduced in preliminary proof-of-concept applications as presented in~\cite{gawron2020-MS-classif-QNN-igarss}, and an ensemble of support vector machines running on the {D-Wave} quantum annealer has been  proposed for remote sensing image classification in~\cite{cavallaro2020-igarss}.  In our preliminary work~\cite{zaidenberg2021advantages_2} hybrid quantum-classical neural networks for remote sensing applications are discussed, and a proof-of-concept for binary classification, using multispectral optical data, is reported. Finally, Otgonbaatar et. al \cite{otgonbaatar2021classification} proposed a binary classifier based on a very deep convolutional network and a 17 qubit quantum circuit.


In this manuscript, different circuit-based hybrid quantum convolutional neural networks (QCNNs) are discussed, and  a remote sensing image classification use case is considered, exceeding the simple binary classification presented in~\cite{zaidenberg2021advantages_2} and the more complex presented in \cite{otgonbaatar2021classification}. Namely, hybrid networks based both on classical and quantum computing will be used, and a comparison will be made of performances  provided, when dealing with different quantum circuits applied to classification of remote sensing images.

The main contributions of this work are as follows:
\begin{itemize}
    \item QC is applied to land-cover classification on the reference benchmark EuroSAT dataset~\cite{ helber2018introducing} for optical multispectral images, thus by going further than initial proofs-of-concept on a few images~\cite{gawron2020-MS-classif-QNN-igarss, cavallaro2020-igarss}.
    \item QCNN multiclass classification is tackled, with respect  to the simple binary classification already discussed in~\cite{zaidenberg2021advantages_2}, and better results are obtained through the quantum-based networks with respect to  their fully-classical counterpart.
    \item A comparative and critical analysis is carried out to analyze the performances of different gate-based circuits for hybrid QCNN, showing the advantages of the architecture with entanglement.
    \item A structured  prediction  setting, with coarse-to-fine classification has been implemented to further challenge the capacities brought by entanglement.
\end{itemize}

Moreover, it is worth to highlight that each model we proposed it has been implemented and designed from scratch. This process involved also the adaptation of the classical and quantum networks to fit the requirements imposed by the used dataset.

It is also worth to mention that this paper can represent a useful tool for machine learning and remote sensing scientists looking at the way quantum circuits and their  parameters work when applied to practical EO problems, since it describes the necessary mathematical and physical elements for the understanding of the quantum approach.
The paper is organized as follows. In Sec.~\ref{sec:lulcoverview} an overview of LULC classification in the field of remote sensing  is given by highlighting  the main issues and difficulties in LULC tasks for remote sensing interpretation. In Sec.~\ref{sec:quantum_ML} the applications of machine learning in the domain of QC are introduced, and in Sec.~\ref{sec:QC_basics} the mathematical and physical background to QC is provided. The proposed methodology and the hybrid QCNNs are presented in Sec.~\ref{sec:methodology}, while the results are reported in Sec.~\ref{sec:results}. Concluding remarks are given in Sec.~\ref{sec:conclusion}. 

\section{Land Use Land Cover Classification Overview }\label{sec:lulcoverview}
LULC classification using remote-sensing imagery has been playing an important role in sustaining, monitoring and planning the usage of natural resources since years. LULC classification has reached a crucial scope in the management of land use, agricultural sector, forest areas and biological resources  \cite{talukdar2020dynamics}, and it has a direct impact on atmosphere, soil erosion and water, while it is indirectly connected to global environmental problems \cite{tsai2019monitoring}, by helping in delivering up-to date and large-scale information on surface conditions.

A general overview of supervised object-based land-cover image classification techniques is reported in \cite{ma2017review}, whereas a more comprehensive and recent review of challenges and state-of-the-art techniques for LULC classification is provided by Talukdar et al. \cite{talukdar2020land}.

For years, classical techniques mainly based on pixel or object analysis in terms of reflectance or local texture have been used for LULC classification  \cite{xiaoxia2005comparison, zarro2020semi}. Yet, they have shown several issues since extremely affected by the data acquisition issues (like cloud cover and regional fog, adaptation to new sensors) and environmental changes which make difficult to design a generic classifier suitable for every object or land class everywhere in the world.

Several new methodologies have been developed by the researchers to address those issues by building on more robust statistical models and in particular the well-known Deep Learning (DL). Two trends have emerged: object-based image analysis (OBIA) or patch-wise classification, and dense pixel-wise classification.

Generally, patch-wise approaches focus on local neighborhoods which correspond to semantically meaningful objects to build the classifiers. The task to achieve is to give a label to a patch which correspond to a small region of a complete aerial or satellite image, as in the popular EuroSAT~\cite{helber2019eurosat} or BigEarthNet~\cite{sumbul2019bigearthnet} benchmarks. Dedicated OBIA methods can then be applied, which look for relevant objet borders for example, as the DOTA baseline which is based on a Region-CNN~\cite{Xia_2018_CVPR}.

On the contrary, pixel-wise approaches follow the historical remote sensing way of modeling local appearance statistics. In the last decade, the use of (Fully-)Convolutional Networks (FCNs) have proved to be extremely efficient by relying on very large models able to capture the diversity of possible inputs, and thus for a large variety of LULC classes: CNNs and random fields~\cite{paisitkriangkrai2015effective}, multi-modal multi-scale FCNs~\cite{audebert2016semantic}, ensemble of CNNs~\cite{marmanis2016semantic}.

Finally, among the new techniques adopted to deal with LULC problems, they must be included strategies based on Capsule networks~\cite{jiang2020hyperspectral}, recurrent networks~\cite{rajendran2020land}, Graph Convolutional Networks (GCNs)~\cite{hong2020graph}, which have be applied to hyperspectral imagery for instance, and  Transformers more recently applied to both patch-wise and pixel-wise classification~\cite{bazi2021vision, xu2021efficient}. Building on this set of powerful tools, new challenges can now be addressed which include explainable and interpretable classification~\cite{tuia2021toward}, weakly-supervised classification~\cite{daudt-weak-CD-ML21}, self-supervised classification, or semi-supervised classification~\cite{castillo-semisup-minifrance-ML21}.\\

After Deep learning, which has proved to be a relevant tool for improving pre-existing classical models, the beginning of the era of quantum computing has brought  new ideas to solve the LULC classification problems, as new opportunities (the amount of data available) but also new issues (large-scale processing, variety of sensors, very high resolution) have appeared.

\section{Quantum Machine Learning}\label{sec:quantum_ML}
As already underlined before, the research presented in this article focuses on the possibility to demonstrate how the use of quantum computers can help in enhancing the performance of ML algorithms when applied to LULC classification.

In this section, a brief review of the recent results and research open questions concerning QML  is first reported. The benefits of QC for ML applications are explained, by highlighting the general advantages of QML and by also presenting some applications. Finally the open challenges of these approaches and existing systems are discussed.  

\textbf{The need for Quantum Computing.} Given the premises of the Introduction section concerning the disruptive potentialities of QC, and the issues discussed in the previous section on the difficulties in LULC tasks for remote sensing interpretation, QML has quickly become a topic of interest for the information science~\cite{DBLP:journals/corr/abs-0705-3360,Biamonte_2017, Abbas_2021, pub.1101409450} since the 1990s. As already anticipated, with the continuously increasing volume of data requiring classification-related processing tasks, computers have had to adapt themselves to process these larger and more complex sets of information. This is why quantum solutions are gaining attention and being explored. Moreover, for ML applications, quantum computers may provide an added benefit since they can avoid getting stuck at relative minima in gradient descent, by quantum tunneling through "hills"~\cite{verdon2019universal}. Practically, quantum computers are likely to reach a better solution than classical computers.  Moreover,  QC provides many other benefits for ML, such as fast linear algebra, quantum sampling, quantum optimization, and quantum artificial neural networks~\cite{Ciliberto2018}. Despite the still unsolved limitations, quantum resources are expected to provide advantages for learning problems.

\textbf{Advantages of Quantum Machine Learning.}As briefly mentioned at the end of  the previous subsection, there are several advantages in using the QC applied to ML, and some examples are found in the literature.
In~\cite{Cong_2019}, for instance, the authors introduce and analyse the QCNN as a machine learning-inspired quantum circuit model, and demonstrate its ability to solve important classes of intrinsically quantum many-body problems. They consider two classes of problems where QC offers some advantages: 1) the quantum phase recognition, which asks whether a given input quantum state belongs to a particular quantum phase of matter, and 2) the quantum error correction (QEC) optimization, where an optimal QEC code is chased, for a given, a priori unknown, error model, such as dephasing or potentially correlated depolarization in realistic experimental settings. 

Currently, different quantum algorithms that could act as building blocks of ML programs have been developed, sometimes related to hardware and software challenges that are not yet completely solved~\cite{Biamonte_2017}.
Given that ML and AI can play fundamental roles in the quantum domain~\cite{pub.1101409450}, the  main benefits of QML, as already summarized in~\cite{phillipson2020}, are the following:
1) improvements in run-time, 2)  learning capacity improvements, 3) learning efficiency improvements. 

However, there is not a shared consensus on how and when QML can be advantageous with respect to its classical counterpart on general classes of problems. For instance, in~\cite{Huang2021}, it is shown how the quality and the amount of data can sensibly affect the performance of classical and QML  models in such a way that the quantum advantage is not always guaranteed. With this regard, this paper adds an important element of discussion with respect to the state of the art, by demonstrating how QML could help when dealing with real remote sensing images for a classification problem where multiple classes are used.

\textbf{Quantum Machine Learning applications.} Currently, there are several general methods for implementing quantum circuits into ML models, as it can be found in the literature. For instance, in~\cite{hernandez2020image}  image  classification is performed via a QML, while in~\cite{Rebentrost_2014}  a quantum support vector machine is used for Big Data classification. In~\cite{henderson2019quanvolutional}  quanvolutional neural networks are employed to carry out image recognition, and instead variational quantum circuits for inductive Grover oracularization are presented in~\cite{hasan2021igoqnn}. Lithology interpretation from well logs is discussed in~\cite{Liu2021}, and quantum variational autoencoder presented in~\cite{Khoshaman2018}. Quantum Neural Networks (QNNs) are often presented as hybrid algorithms that leverage quantum nodes throughout the networks~\cite{article123,Liu2021Hybrid,Oh2020}. QNNs develop a network of both quantum and classical nodes with some given activation functions, convolutional connections, and weighted edges. Here, the quantum nodes can be represented by single qubits or clusters of qubits. QNNs can also present a more complexly integrated circuit with entanglement, where correlations between quantum nodes can be exploited to speed up computation. 

\textbf{Quantum Machine Learning challenges.} Trying to create  complex quantum networks which link together layers of quantum nodes still represents a research challenge. 
Despite the many possible theoretical applications of quantum computers, there is still significant progress that must be made towards more reliable computation. The QC industry currently finds itself in the Noisy Intermediate-Scale Quantum (NISQ) era, where there is a limit to the number of operations that can be performed on a quantum computer before the information stored becomes useless~\cite{bharti2021noisy}. Currently, these limitations contribute to the difficulties in scaling up quantum computers. However, all the work in progress is not useless since as soon as scaling quantum computers become viable, they will be able to represent exponentially more information than the classical ones. 
Fortunately, recent events show promising evidence  for moving ahead and away from the NISQ era. In particular, by using QCNN models, researchers have been able to create an optimal QEC scheme for a given error mode~\cite{Cong_2019}, and moreover, many QC  companies are also projecting similar timelines for developing their architecture. Some companies are planning to release error corrected and fault tolerant commercial quantum computers by the 2025~\cite{psiQuantum,Goldman_Sachs}.


\section{Mathematical Background on QC}
\label{sec:QC_basics}

In this section the basic notions of quantum computing are introduced. Further information can be retrieved in~\cite{Kaye_intro_quantum,nielsen2011}.

\begin{figure}[!ht]
\centering
\resizebox{0.35\columnwidth}{!}{
\begin{tikzpicture}[ line cap=round, line join=round, >=Triangle]
  \draw[dash pattern=on 5pt off 3pt] (0,0) circle (2cm);
  \draw[dash pattern=on 5pt off 3pt] (0,0) circle (1cm);
  \draw (0.85,0.85) node[anchor=north west] {$\ket{0}$};
  \draw (1.6,1.6) node[anchor=north west] {$\ket{1}$};
  \draw [fill] (0,0) circle (3pt);
\end{tikzpicture}}
\caption{Qubit modeling as hydrogen atom, with electron ground state $\ket{0}$ and first exited state $\ket{1}$.} \label{fig:H_qubit}
\end{figure}
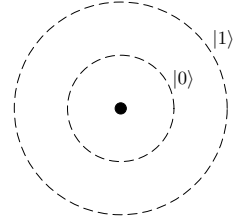

\textbf{Qubits} are the fundamental units of information held in quantum computers. 
A physical qubit exists in a \emph{superposition} of two states, $\ket{0}$ and $\ket{1}$, as shown in Fig.~\ref{fig:H_qubit} referring to an hydrogen atom with ground and exited states. The state $\ket{\psi}$ of the qubit describes the probability distribution of the state and is expressed as
\begin{equation}
    \centering
    \ket{\psi} = \alpha\ket{0} + \beta\ket{1}.
    \label{eqn:state_qubit}
\end{equation}

\textbf{Quantum measurement} is an irreversible operation in which information is gained about the state of a single qubit, and superposition is lost.
Mathematically speaking, in Eq. \eqref{eqn:state_qubit} $\ket{\psi}$ can be viewed as a vector in a Hilbert Space (i.e., a vector space equipped with an inner product operation) where%
\begin{equation}
    \centering
    \ket{0} = 
        \begin{pmatrix}
          1\\ 
          0
        \end{pmatrix},
        \quad
        \ket{1} = 
    \begin{pmatrix}
      0\\ 
      1
    \end{pmatrix},
\end{equation}
$\alpha,\,\beta\in\mathbb{C}$ represent the probability of measuring the state $\ket{0}$ and $\ket{1}$, respectively, with the constraint $\abs{\alpha^2}+\abs{\beta^2}=1$. For the state $\ket{\psi} = \sqrt{\frac{1}{3}}\ket{0}  + \sqrt{\frac{2}{3}}\ket{1}$, the probabilities of measuring $\ket{0}$ and $\ket{1}$ are $\frac{1}{3}$ and $\frac{2}{3}$, respectively. Moreover, the measurement process does irreversibly modify the qubit, so that after the measurement the qubit can be $\ket{\psi}=\ket{0}$ with probability $\alpha^2$, and $\ket{\psi}=\ket{1}$ with probability $\beta^2$.

When considering a system of two qubits with states $\alpha_0\ket{0} + \alpha_1\ket{1}$ and $\beta_0\ket{0} + \beta_1\ket{1}$, the state evaluated by means of the tensor product is the superposition given by 
\begin{equation}
    \ket{\psi} = \alpha_{0}\beta_{0}\ket{00}
    +\alpha_{0}\beta_{1}\ket{01}
    +\alpha_{1}\beta_{0}\ket{10}
    +\alpha_{1}\beta_{1}\ket{11},
\label{eq:entanglement}
\end{equation}
where  $\alpha_{i},\beta_{j}\in\mathbb C$ and $\sum\alpha_i\beta_j=1$. The state $\ket{00}$, for instance, is given as $\ket{0}\otimes{\ket{0}}$, where $\otimes$ is the tensor product. It turns out that, in general, you cannot factorize the state in Eq. \eqref{eq:entanglement} in terms of the original qubits. This phenomenon, known as \emph{entanglement}, has an important consequence in the measurement process. Indeed, considering the \emph{Bell} state
\begin{equation}
    \ket{\psi} = \frac{1}{\sqrt{2}}\ket{00}
    +\frac{1}{\sqrt{2}}\ket{11},
\label{eq:Bell}
\end{equation}
if the measurement of the first qubit returns the state $\ket{0}$ (with probability $0.5$), than the entangled state collapse to $\ket{00}$. At this point, the second qubit is completely known as it is in the state $\ket{0}$ as well. This result is true even when the two qubits are separated by a very large (theoretically infinite) distance, leading to the violation of the locality principle of classical mechanics. By using the Schmidt decomposition theorem, it can be shown that a quantum system can have different degrees of entanglement \cite{nielsen_chuang_2010}. By exploiting superposition and entanglement, quantum computers can perform operations that are difficult to emulate on a large scale with classical computers, cutting down computational time and power to process information. 

\begin{figure}[!ht]
\centering
\resizebox{0.4\columnwidth}{!}{
\begin{tikzpicture}[line cap=round, line join=round, >=Triangle]
  \clip(-2.19,-2.49) rectangle (2.66,2.58);
  \draw [shift={(0,0)}, lightgray, fill, fill opacity=0.1] (0,0) -- (56.7:0.4) arc (56.7:90.:0.4) -- cycle;
  \draw [shift={(0,0)}, lightgray, fill, fill opacity=0.1] (0,0) -- (-135.7:0.4) arc (-135.7:-33.2:0.4) -- cycle;
  \draw(0,0) circle (2cm);
  \draw [rotate around={0.:(0.,0.)},dash pattern=on 3pt off 3pt] (0,0) ellipse (2cm and 0.9cm);
  \draw (0,0)-- (0.70,1.07);
  \draw [->] (0,0) -- (0,2);
  \draw [->,dash pattern=on 3pt off 3pt] (0,0) -- (0,-2);
  \draw [->] (0,0) -- (-0.81,-0.79);
  \draw [->] (0,0) -- (2,0);
  \draw [dotted] (0.7,1)-- (0.7,-0.46);
  \draw [dotted] (0,0)-- (0.7,-0.46);
  \draw (-0.08,-0.3) node[anchor=north west] {$\varphi$};
  \draw (0.01,0.9) node[anchor=north west] {$\vartheta$};
  \draw (-1.01,-0.72) node[anchor=north west] {$\mathbf {\hat{x}}$};
  \draw (2.07,0.3) node[anchor=north west] {$\mathbf {\hat{y}}$};
  \draw (-0.5,2.6) node[anchor=north west] {$\mathbf {\hat{z}=\ket{0}}$};
  \draw (-0.4,-2) node[anchor=north west] {$-\mathbf {\hat{z}=\ket{1}}$};
  \draw (0.4,1.65) node[anchor=north west] {$|\psi\rangle$};
  \scriptsize
  \draw [fill] (0,0) circle (1.5pt);
  \draw [fill] (0.7,1.1) circle (0.5pt);
\end{tikzpicture}}
\caption{The Bloch sphere representing the probabilistic space in which the quantum state can exist. Gate operations rotate $\ket{\psi}$ about the Bloch sphere, changing the phase and the probability amplitudes of the qubit.} 
\label{fig:bloch_sphere}
\end{figure}
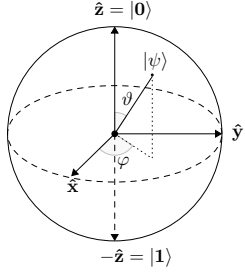

The qubit state in Eq. \eqref{eqn:state_qubit} can be expressed as a function of two angles $\vartheta$ and $\varphi$, i.e.
\begin{equation}
    \ket{\psi} = \cos{\frac{\vartheta}{2}}\ket{0} + e^{i\varphi}\sin{\frac{\vartheta}{2}}\ket{1},
\end{equation}
and represented as a point
sitting on the surface of a unitary three-dimensional sphere, named the Bloch sphere, as shown in Fig.~\ref{fig:bloch_sphere}. With this notation, $\vartheta$ describes the probability of the qubit to result in $\ket{0}$ or $\ket{1}$ and the angle $\varphi$ describes the phase the qubit is in.

\textbf{Quantum gates}, denoted by $U$ in the following, are basic quantum circuits operating on a small number of qubits. They are the building blocks of quantum circuits, like classical logic gates are for conventional digital circuits. Quantum gates are unitary operators, i.e. $U^{\dagger} U = U U^{\dagger} = I$, where the symbol $\dagger$ denotes the conjugate transpose, and $U$ is described as a unitary matrix relative to some basis. Important properties are that 1) $U$ preserves the inner product of the Hilbert space and 2) qubit gate operations can also be visualized as rotations of the quantum state vector in the Bloch sphere. \\

The standard quantum gates used in this paper are introduced hereafter:
\begin{itemize}
    \item \textit{Hadamard} gate, a single qubit gate described by the matrix: 
\begin{equation}
    \centering
    H = \frac{1}{\sqrt{2}}
        \begin{pmatrix}
          1 & 1\\ 
          1 & -1
        \end{pmatrix}.
\end{equation}
Starting from the single state qubit $\ket{0}$, the Hadamard gate return the superposition of two states, namely the so called \emph{plus} state $\ket{+}$, i.e.
\begin{equation}
\begin{aligned}
    \centering
    H \ket{0} &= \frac{1}{\sqrt{2}}
        \begin{pmatrix}
          1 & 1\\ 
          1 & -1
        \end{pmatrix}
        \begin{pmatrix}
          1\\ 
          0
        \end{pmatrix} = \frac{1}{\sqrt{2}}
        \begin{pmatrix}
          1\\ 
          1
        \end{pmatrix}\\
        & =\frac{1}{\sqrt{2}}
        \begin{pmatrix}
          1\\ 
          0
        \end{pmatrix}
        + \frac{1}{\sqrt{2}}
        \begin{pmatrix}
          0\\ 
          1
        \end{pmatrix}\\
        &= \frac{1}{\sqrt{2}}
        \ket{0}
        + \frac{1}{\sqrt{2}}
        \ket{1} = \ket{+}
\end{aligned}
\label{eq:hadamard}
\end{equation}

    \item \textit{Rotation} gates, $ R_{x}(\theta )$, $R_{y}(\theta )$, $R_{z}(\theta )$, i.e. single qubit gates described by rotation matrices about the $\mathbf {\hat{x}}$, $\mathbf {\hat{y}}$, $\mathbf {\hat{z}}$ axes of the Bloch sphere, respectively. The gate $R_{y}(\theta )$, which will be used in the following, takes the form:
    \begin{equation}
        R_{y}(\theta ) = 
        \begin{pmatrix}
            \cos{\frac{\theta}{2}} & -\sin{\frac{\theta}{2}}\\
            \sin{\frac{\theta}{2}} & \cos{\frac{\theta}{2}}
        \end{pmatrix}.
    \end{equation}
    
    \item \textit{CNOT} gate, which is a two qubits gate described by the matrix
    \begin{equation}
        U = \begin{pmatrix}
          1&0&0&0\\
          0&1&0&0\\
          0&0&0&1\\
          0&0&1&0
        \end{pmatrix}
    \end{equation}
    and represented in Fig.~\ref{fig:cnot_gate}. When the input are basis states $\ket{0}$ and $\ket{1}$, the CNOT gate transform the state
    \begin{equation*}
     \alpha_{00}\ket{00}
    +\alpha_{01}\ket{01}
    +\alpha_{10}\ket{10}
    +\alpha_{11}\ket{11}
    \end{equation*}
    into 
    \begin{equation*}
     \alpha_{00}\ket{00}
    +\alpha_{01}\ket{01}
    +\alpha_{10}\ket{11}
    +\alpha_{11}\ket{10},
    \end{equation*}
    i.e., it flips the second qubit (the target qubit) if and only if the first qubit (the control qubit) is $\ket{1}$.\par
\begin{figure}[!ht]
\centering
\resizebox{0.35\columnwidth}{!}{
\begin{quantikz}
\lstick{$\ket{\psi_1}$} & \ctrl{1} & \meter{}\\
\lstick{$\ket{\psi_2}$} & \targ{} & \meter{}\\
\end{quantikz}}
\caption{CNOT gate with two input qubits and measurement output.}
\label{fig:cnot_gate}
\end{figure}
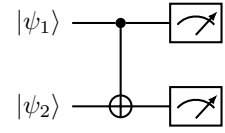
\end{itemize}

\begin{figure}[!ht]
\centering
\resizebox{0.4\columnwidth}{!}{
\begin{quantikz}
\lstick{$\ket{0}$}&\gate{H} & \ctrl{1}& \qw \\
\lstick{$\ket{0}$} & \qw & \targ{}& \qw \\
\end{quantikz}}
\caption{Quantum circuit to create Bell state.}
\label{fig:Bell_circuit}
\end{figure}
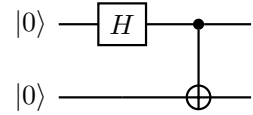

The combination of Hadamard and CNOT gates is used to create an entangled Bell state as defined in Eq. \eqref{eq:Bell}. The corresponding circuit shown in Fig.~\ref{fig:Bell_circuit} is the basic building block of the quantum circuits investigated in this paper, as it introduces entanglement in the circuit by enhancing the computation performances. 

\section{Methodology}\label{sec:methodology}
In this section, a selected number of quantum circuits, investigated as potential quantum layers in the proposed hybrid network, are described. Firstly, the integration of the quantum part into the classical architecture is discussed, by presenting the \textit{"Data Embedding"} operation and showing an example of interface between classical and quantum layers. At the end of the section, the hybrid QCNN is presented, and the model optimization and inference discussed. Although the quantum circuits presented in the following are standardly used in QC for data processing and they are fundamental units of IBM Qiskit~\cite{asfaw2020learn, githubquantumdl}, it is worth to highlight that all codes have been realized from scratch by the authors and released open-access in a public repository~\cite{githubrepo}.

\subsection{Data Embedding}
To  create  a  hybrid  QNN,  a  parametrized  quantum  circuit is typically used as a hidden layer for the neural network. Yet, with respect to classical network architectures, right in order to integrate the quantum part into the classical architecture, it is critical to realize a higher dimensional quantum representation of classical data in the creation of the hybrid model. In this section, a brief description on how to prepare a quantum state at this end is given. 

A feature mapping is first run through a unitary operator applied to a set of N $\ket{0}$ quantum nodes as a method of encoding the classical information in the new N-qubit space. A unitary matrix, needed to encode the information, must be classically derived before applying it to the quantum circuit. Its parameters are determined by the values of the preceding classical nodes at the point of insertion. This operation is referred to as data embedding, where the preceding classical activation is represented through the related amplitude probability of measuring $\ket{1}$ in the quantum state. 

Different gate operations can be used to encode a quantum representation of classical information. For instance, Abbas et al. in~\cite{Abbas_2021} show how that can be done by first applying a Hadamard gate to put the qubits in a superposition state, and then  by applying RZ-gate rotations to the qubits, with angles equivalent to the feature values of preceding inputs. Alternate gate operations can be used to encode a quantum representation of classical information. Yet, the interpretation of the prepared state must be self consistent, that means to consider the encoding system valid as long as the input operations and the output measurement accurately represent the classical information.

\begin{figure}
    \centering
    \resizebox{0.85\columnwidth}{!}{
    \begin{tikzpicture}[node distance=2cm]
    
    \node (x2) [inout] {\Large $x_2$};
    \node (x1) [inout, left of = x2, xshift = -0.3\columnwidth] {\Large $x_1$};
    \node (x3) [inout, right of = x2, xshift = 0.3\columnwidth] {\Large $x_3$};
    \node (init) [textbox, left of = x1, text width=2cm, align=left] {\Large  \textit{Classical\\network\\(input)}};
    
    \node (h1) [hidden, below of=x2, xshift = -0.4\columnwidth] {\Large $h_1$};
    \node (h2) [hidden, below of=x2, xshift =  0.4\columnwidth] {\Large $h_2$};

    \draw [arrow] (x1) -- node[near end,left] {\Large $\omega_1$} (h1);
    \draw [arrow] (x2) -- node[near end,above] {\Large $\omega_2\;\;$} (h1);    
    \draw [arrow] (x3) -- node[near end,below] {\Large $\omega_3$} (h1);
    
    \draw [arrow] (x1) -- node[near end,below] {\Large $\omega_4$} (h2);
    \draw [arrow] (x2) -- node[near end,above] {\Large $\omega_5$} (h2);
    \draw [arrow] (x3) -- node[near end,right] {\Large $\Large \omega_6$} (h2);
    
    \node (h1t) [textbox, below of=h1] {\Large $h_1=\sigma(\omega_1 x_1+\omega_2 x_2+\omega_3 x_3)$};
    \node (h2t) [textbox, below of=h2] {\Large $h_2=\sigma(\omega_4 x_1+\omega_5 x_2+\omega_6 x_3)$};
    
    \draw [arrow] (h1) -- (h1t);
    \draw [arrow] (h2) -- (h2t); 
    
    \node (init2) [textbox, below of = h2t, xshift = -1.15\columnwidth,, text width=2cm, align=left] {\Large \textit{Quantum\\circuit}};
    
	
	\node (r1) [gate, below of=init2, xshift = 0.35\columnwidth,minimum size=1.6cm,draw] {\Large $\!\!R(h_1)$};
	\node (m1) [met, right of = r1, xshift = 0.8\columnwidth,minimum size=1.2cm] {\Large $\!m_1$};
	\node (psi1) [textbox, left of= r1, xshift = -0.1\columnwidth] {\Large $\ket{\Psi_1}$};	
	\draw [arrow] (h1t) -- (r1);
	\draw [arrow] (psi1) -- (r1);
	\draw [arrow] (r1) -- (m1);	
	
	\node (psi2) [textbox, below of=psi1] {\Large $\ket{\Psi_2}$};	
	\node (r2) [gate, right of=psi2, xshift = 0.9\columnwidth,minimum size=1.6cm,draw] {\Large$\!\!R(h_2)$};
	\node (m2) [met, below of = m1,minimum size=1.2cm] {\Large $\!m_2$};
	
	\draw [arrow] (h2t) -- (r2);
	\draw [arrow] (psi2) -- (r2);
	\draw [arrow] (r2) -- (m2);

    \node (h3) [hidden, below of=m2, xshift = -1.03\columnwidth] {\Large $h_3$};
    \node (h4) [hidden, below of=m2, xshift =  -0.23\columnwidth] {\Large $h_4$};
    \node (init3) [textbox, left of = h3, xshift = -0.1\columnwidth, text width=2cm, align=left] {\Large \textit{Classical\\network\\(output)}};
    
    \draw [arrow] (m1) -- (h3);
    \draw [arrow] (m2) -- (h4); 
    
    \node (y) [inout, below of = h3, xshift = 0.4\columnwidth] {\Large $y$};
    
    \draw [arrow] (h3) -- node[near end,left] {\Large $\omega_7\;\;$}(y);
    \draw [arrow] (h4) -- node[near end,right] {\Large $\;\;\omega_8$}(y); 
    
    \node (yt) [textbox, right of=y, xshift = 0.2\columnwidth] {\Large $y=\sigma(\omega_7 h_3+\omega_8 h_4)$};
    \end{tikzpicture}
}
    \caption{Interface between classical and quantum layers.}
    \label{fig:interface}
\end{figure}
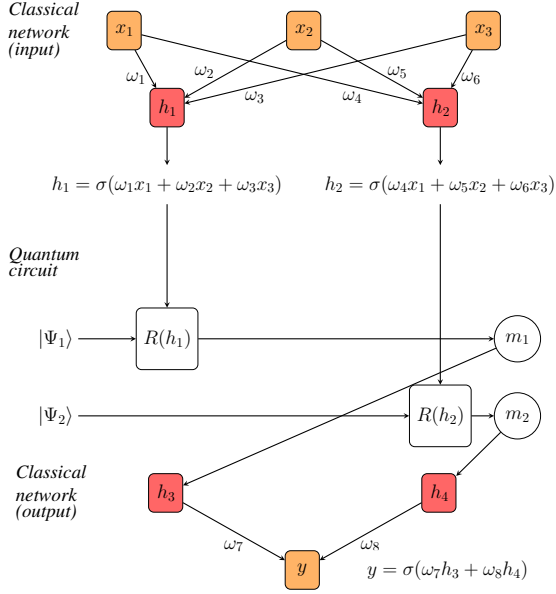

Proceeding the classical encoding, the parametrized quantum circuit is then applied. A parametrized quantum circuit is a quantum circuit where the rotation angles for each gate are specified by the components of a classical input vector. The outputs from the neural network's previous layer will be collected and used as the inputs for the parametrized circuit. The measurement statistics of the quantum circuit can then be collected and used as inputs for the following hidden layer. As a demonstrative example in Figure~\ref{fig:interface} the interface between classical and quantum layers is sketched.

\subsection{Selected Quantum Circuits for Image Classification}
Three types of circuits, selected among the possible quan\nobreak tum circuits and to be used in the proposed hybrid QCNN, are presented. Their structure reflects the adopted implementation with 4 qubits, which represents a more complex architecture with respect to simpler ones where less qubits are used~\cite{zaidenberg2021advantages_2}. Far from being an exhaustive comparison of all possible quantum configurations, the description of the adopted circuits will allow to get an insight on how their gates can influence the final results and help speed up certain computational processes. To better understand how the entangled qubits, introduced in Sec. \ref{sec:QC_basics}, can affect the classification performance, it is necessary to clarify that the first circuit has no entanglement, whereas entanglement is introduced in the remaining ones through different gate connections.

\begin{figure}[!ht]
\centering
\resizebox{0.5\columnwidth}{!}{
\begin{quantikz}
    \lstick{$\ket{\psi_0}$} & \gate{H} & \gate{R_y(\theta_0)} & \meter{}\\
    \lstick{$\ket{\psi_1}$} & \gate{H} & \gate{R_y(\theta_1)} & \meter{}\\
    \lstick{$\ket{\psi_2}$} & \gate{H} & \gate{R_y(\theta_2)} & \meter{}\\
    \lstick{$\ket{\psi_3}$} & \gate{H} & \gate{R_y(\theta_3)} & \meter{}
\end{quantikz}}
\caption{No entanglement circuit.}
\label{fig:circuit1}
\end{figure}
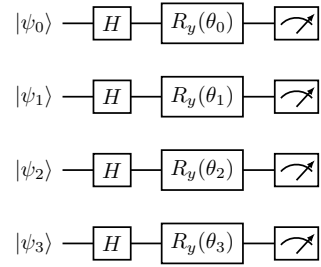

\textbf{No entanglement circuit}
In the simple QCNN presented in~\cite{zaidenberg2021advantages_2}, there is no entanglement and classical nodes are merely replaced by a parameters quantum node~\cite{article123}. As seen in Fig.~\ref{fig:circuit1}, the qubits are first placed in superposition through the application of a Hadamard gate.  

Next, the quantum nodes undergo $R_y$ gate rotations about the parameters $\theta$. This whole process is ultimately representative of quantum node activation which simply encodes the sum of the weighted activations from preceding classical nodes that are mapped into the quantum nodes. If only one qubit is considered, the effect of the Hadamard and rotation gates on the qubit $\ket{0}$ are summarized as:

\begin{equation}
\begin{aligned}
    \centering
    R_y(\theta)H \ket{0}&=\frac{1}{\sqrt{2}}\begin{pmatrix}
     \cos(\frac{\theta}{2}) & -\sin(\frac{\theta}{2})\\ 
     \sin(\frac{\theta}{2}) & \cos(\frac{\theta}{2})
        \end{pmatrix}
        \begin{pmatrix}
          1 & 1\\ 
          1 & -1
        \end{pmatrix}
        \begin{pmatrix}
          1 \\ 
          0
        \end{pmatrix}\\
    &\stackrel{\text{Eq. \eqref{eq:hadamard}}}{=
    }\frac{1}{\sqrt{2}}
       \begin{pmatrix}
     \cos(\frac{\theta}{2}) & -\sin(\frac{\theta}{2})\\ 
     \sin(\frac{\theta}{2}) & \cos(\frac{\theta}{2})
        \end{pmatrix} \left(
        \ket{0}
        +    \ket{1}\right)\\
    &=\frac{\cos(\frac{\theta}{2})+\sin(\frac{\theta}{2})}{\sqrt{2}}\ket{0} + \frac{\cos(\frac{\theta}{2})-\sin(\frac{\theta}{2})}{\sqrt{2}}\ket{1}
\end{aligned}
\end{equation}

The overall gate composed by 4 Hadamard and 4 rotation gates can be built by using the matrix multiplication for successive gates and the tensor product for parallel gates, hence the final unitary transformation $U$ is
\begin{equation}
 U^* =\bigotimes_{i=0}^{i=3}
 \left(R_y(\theta_i) \cdot H\right)
\end{equation}

The entire circuit returns the state

\begin{equation}
    \ket{\psi} = U^*(\ket{\psi_0}\otimes
    \ket{\psi_1}\otimes
    \ket{\psi_2}\otimes
    \ket{\psi_3})
\end{equation}
which, when considering $\ket{0}$ as inputs, is
\begin{equation}
    \ket{\psi} = U^*\ket{0000}.
\end{equation}

\begin{figure}[!ht]
\centering
\resizebox{0.9\columnwidth}{!}{
\begin{quantikz}
\lstick{$\ket{\psi_0}$}&\gate{H}&\ctrl{1}&\qw     &\qw     &\gate{R_y(\theta_0)}&\qw     &\qw     &\ctrl{1}&\qw&\meter{}\\
\lstick{$\ket{\psi_1}$}&\qw     &\targ{} &\ctrl{1}&\qw     &\gate{R_y(\theta_1)}&\qw     &\ctrl{1}&\targ{} &\qw&\meter{}\\
\lstick{$\ket{\psi_2}$}&\qw     &\qw     &\targ{} &\ctrl{1}&\gate{R_y(\theta_2)}&\ctrl{1}&\targ{} &\qw     &\qw&\meter{}\\
\lstick{$\ket{\psi_3}$}&\qw     &\qw     &\qw     &\targ{} &\gate{R_y(\theta_3)}&\targ{} &\qw     &\qw     &\qw&\meter{}
\end{quantikz}}
\caption{Bellman circuit.}
\label{fig:bellman}
\end{figure}
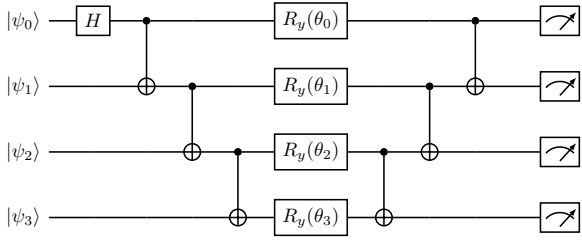

\textbf{Bellman Circuit}
The Bellman Circuit shown in Fig.~\ref{fig:bellman} leverages a basic system of entanglement to encode classical information into a quantum space. Here the speedup may lie in the fact that the quantum states are prepared first through entanglement (by means of the Hadamard and CNOT gates) leading to correlational associations. Following the entanglement process, the parametrization using angular rotations predefined by classical information once more translate the classical information as a quantum activation. 

The qubits are first entangled through the application of a Hadamard gate and then sequential CNOT gates. Following this, the qubits are rotated about the y axis using parameters $\theta$. This is the basis of the activation process. Then the CNOT application process is reversed, but the superposition is never removed. The benefit of this process seems to lie in the variation of the encoding and rotation process, as it is now not just a projection of the classical information into a quantum space, but rather a transformation of this information that exploits quantum feature space. 

Considering the four inputs as $\ket{0}$, before entering into the rotation gates the state of the 4 qubits is given as
\begin{equation}
    \ket{\psi} = \frac{1}{\sqrt{2}}\ket{0000} + \frac{1}{\sqrt{2}}\ket{1111}.
\end{equation}
The four rotation gates applied to the entangled state correspond to the application of the gate
\begin{equation}
    R_y^{\otimes4}(\theta_0,\theta_1,\theta_2,\theta_3) = 
    \bigotimes_{i=0}^{i=3}
    R_y(\theta_i)
\label{eq:R_tensor4}
\end{equation}
corresponding to a $16\times16$ matrix. Finally, the rotated entangled state passes through three more CNOT gates and then it is measured. Supposing the four rotations are identities (i.e., $\theta_i=0,\, i=1,\ldots,4$), the effect of the three CNOT gates is
\begin{equation*}
\begin{gathered}
    \frac{1}{\sqrt{2}}\ket{0000} + \frac{1}{\sqrt{2}}\ket{1111}
    \xrightarrow{1^{st}\,\textrm{CNOT}}
    \frac{1}{\sqrt{2}}\ket{0000} + \frac{1}{\sqrt{2}}\ket{1110},\\
    \frac{1}{\sqrt{2}}\ket{0000} + \frac{1}{\sqrt{2}}\ket{1110}
    \xrightarrow{2^{nd}\,\textrm{CNOT}}
    \frac{1}{\sqrt{2}}\ket{0000} + \frac{1}{\sqrt{2}}\ket{1100},\\
    \frac{1}{\sqrt{2}}\ket{0000} + \frac{1}{\sqrt{2}}\ket{1100}
    \xrightarrow{3^{rd}\,\textrm{CNOT}}
    \frac{1}{\sqrt{2}}\ket{0000} + \frac{1}{\sqrt{2}}\ket{1000}.
\end{gathered}    
\end{equation*}

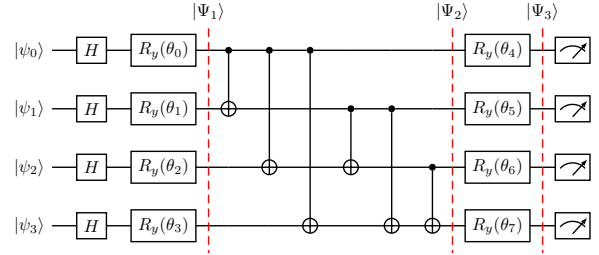
\begin{figure}[!ht]
\centering
\resizebox{0.9\columnwidth}{!}{
\begin{quantikz}
\lstick{$\ket{\psi_0}$}&\gate{H}&\gate{R_y(\theta_0)}\slice{$\ket{\Psi_1}$} &\ctrl{1}&\ctrl{2}&\ctrl{3}&\qw     &\qw     &\qw  \slice{$\ket{\Psi_2}$}   &\gate{R_y(\theta_4)}\slice{$\ket{\Psi_3}$}&\meter{}\\
\lstick{$\ket{\psi_1}$}&\gate{H}&\gate{R_y(\theta_1)}&\targ{} &\qw     &\qw     &\ctrl{1}&\ctrl{2}&\qw     &\gate{R_y(\theta_5)}&\meter{}\\
\lstick{$\ket{\psi_2}$}&\gate{H}&\gate{R_y(\theta_2)}&\qw     &\targ{} &\qw     &\targ{} &\qw     &\ctrl{1}&\gate{R_y(\theta_6)}&\meter{}\\
\lstick{$\ket{\psi_3}$}&\gate{H}&\gate{R_y(\theta_3)}&\qw     &\qw     &\targ{} &\qw     &\targ{} &\targ{} &\gate{R_y(\theta_7)}&\meter{}
\end{quantikz}}
\caption{Real Amplitudes circuit.}
\label{fig:real_amplitude}
\end{figure}

\textbf{Real Amplitudes Circuit}
As is shown in Fig.~\ref{fig:real_amplitude}, breaking down the circuit, each qubit passes through a Hadamard gate and then undergoes a gate rotation with parameters $\theta$ (this value is derived from the result of the preceding classical node). This is the process by which the classical information is turned into quantum information. Then, the qubits are all mutually entangled using CNOT gates.
For instance, considering identity rotations, i.e. $R_y(\theta_i)=I,\,i=0,\ldots,3$, the state before the CNOT gates is
\begin{equation*}
\begin{aligned}
\ket{\Psi_1} &= \left(\bigotimes_{i=0}^{i=3}
 H\right)\ket{0000} = \bigotimes_{i=0}^{i=3}
 \left(H\ket{0}\right) \\
 &= \left(\frac{1}{\sqrt{2}}\right)^4\bigotimes_{i=0}^{i=3}
 \left(\ket{0} + \ket{1}\right) \\
& = 0.25 (\ket{0000} + \ket{0010} + \ket{0011} + \ket{0001} \\ 
& +\ket{0100} + \ket{0110} + \ket{0111} +
\ket{0101}  \\
& +\ket{1000} + \ket{1010} + \ket{1011} + \ket{1001} \\
& +\ket{1100} + \ket{1110} + \ket{1101} + 
\ket{1111} ) .
\end{aligned}
\end{equation*}
After the CNOT gates, one can easily verify that this example state is unchanged, i.e. $\ket{\Psi_1}=\ket{\Psi_2}$ (but in the general case it varies). Finally, the quantum parameters $\theta_i,\,i=4,\ldots,7$ are implemented by means of the final four rotations. By using Eq. \eqref{eq:R_tensor4}, the final state is
\begin{equation*}
    \ket{\Psi_3} = R_y^{\otimes4}\ket{\Psi_2}.
\end{equation*}
During the validation and testing process, the second $\theta$ parameters are used as the ``quantum weights" mapping to the following classically fully connected layer of the nodes.

\subsection{Hybrid Quantum Neural Network Classifier}
Differently from fully Quantum AI models, the proposed QCNN classifier is based on recent hybrid QML models and it consists of the combination of classical ML and quantum layers~\cite{verdon2019universal, schuld2014quest}. This kind of paradigm~\cite{liang2021hybrid, Mari2020transferlearning}, mostly used in the computer vision domain, in this paper has been transferred and adapted to the Remote Sensing domain. Moreover, it is worth highlighting that the hybrid solutions are the preferred ones in the current stage of QML, mostly due to technology bottlenecks and limitations~\cite{zaidenberg2021advantages_2, otgonbaatar2021classification}.

\begin{figure*}[!ht]
    \centering
    \includegraphics[width=2\columnwidth]{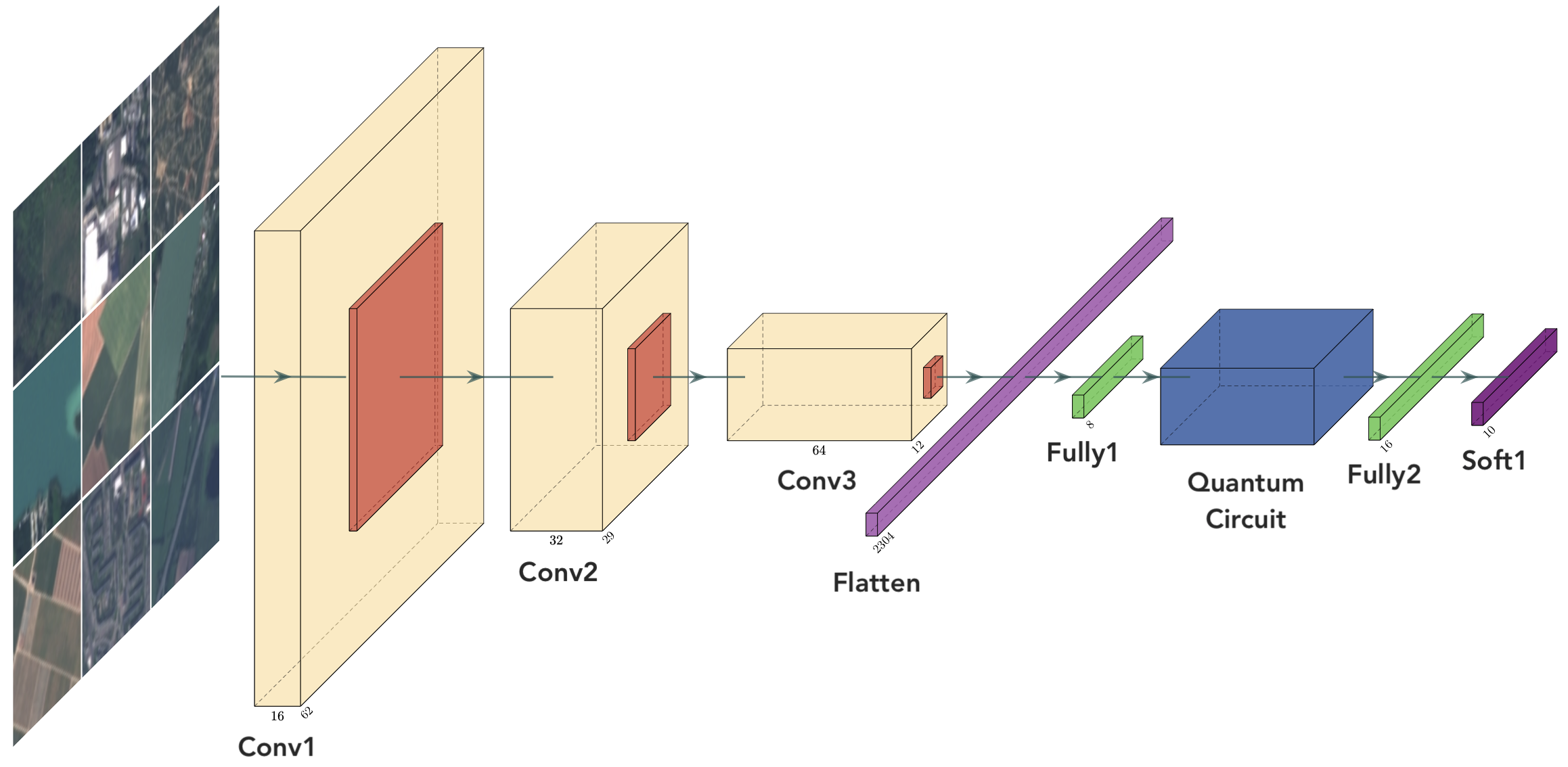}
    \caption{Proposed hybrid Quantum Neural Network Classifier \footnotemark. The network is a modified version of LeNet-5, where the blue box indicates the Quantum Circuit layer.}
    \label{fig:model}
\end{figure*}

The Fig.~\ref{fig:model} shows the QCNN structure, where the classical part consists of a CNN derived from the LeNet-5~\cite{lecun2015lenet}, in which both the number of convolutional layers and the input dimension were changed to fit the input image size. Moreover, with respect to the original LeNet-5 design, the proposed model contains only two fully connected layers, stacked before and after the quantum layer. These two layers are used respectively for adapting the input size, needed by the quantum layer, and the quantum layer output size to match the number of classes imposed by the chosen dataset. In other words, the purpose of these two classical neural layers is to ensure \emph{data embedding} from the image space to the quantum capacity and to make possible the coexistence of classical and quantum layers in the hybrid structure.

Regarding the quantum part, the quantum layer (blue box labeled as Quantum Circuit in Fig.~\ref{fig:model}) aims to benefit of the properties of probabilistic quantum computing. This quantum layer is implemented with one the circuits described in Sec.~\ref{sec:methodology}. In the course of this study, several quantum circuits were tested and analyzed to investigate their potential. 

For comparisons purposes, two versions of the classical counterpart of the proposed QCNN classifier have been implemented and tested.  For the classical CNN classifier 1, the quantum circuit has been replaced with a fully connected layer of 16 nodes, based on the quantum circuit output size. For the classical CNN classifier 2, the quantum circuit has been replaced with a multi-layer perceptron with fully-connected layers of 256, 64, 32, 10 nodes.

\begin{figure}[!ht]
    \centering
    \includegraphics[width=1\columnwidth]{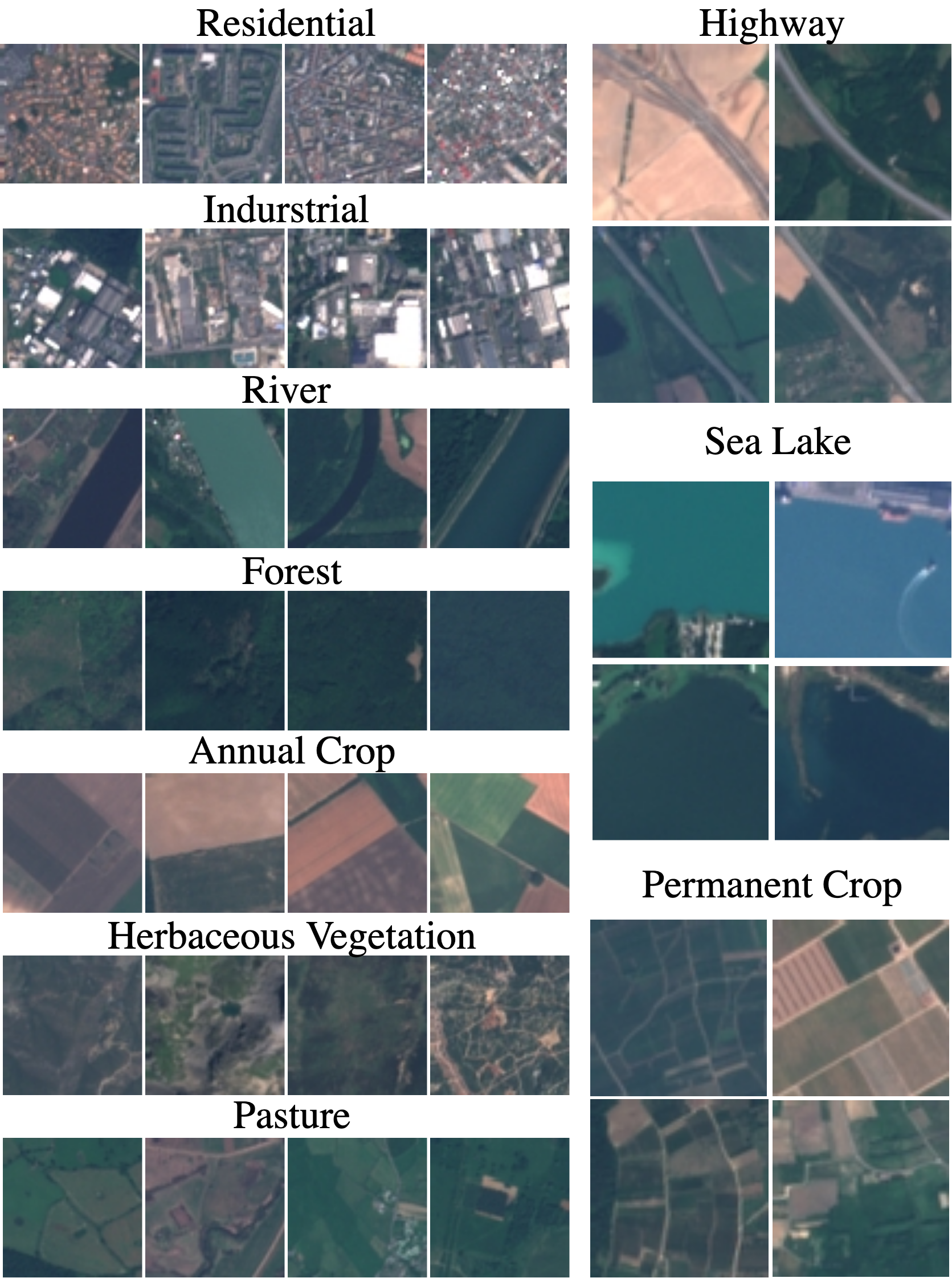}
    \caption{Sample of EuroSAT dataset, 4 images for each class. }
    \label{fig:datasetsample}
\end{figure}

The experimental dataset under consideration is the "EuroSAT: Land Use and Land Cover Classification with Sentinel-2", a dataset of Sentinel-2 satellite images covering 13 spectral bands and consisting out of 10 classes with in total 27.000 labeled and geo-referenced images~\cite{helber2019eurosat}. The dataset has been divided in training and validation sets with a 80-20 factor. Sample images of the dataset are shown in Fig.~\ref{fig:datasetsample}. 

In the following sections several experiments have been carried out, such as: 1) experiments on 3 different quantum circuits, 2) experiments on 2 classical deep learning models for comparison with the quantum counterpart, 3) experiments on a coarse quantum classifier and 3 fine-grain quantum classifiers and 4) an additional experiment, involving the fine-grain classifier, to create a segmentation map.

As highlighted at the beginning of this section, it is fair to remark that all the proposed models were implemented and designed from scratch. This process involved also the adaptation of the classical and quantum networks to fit the requirements imposed by the dataset used for the experimental analysis. No pre-trained weights were used and also the selection of hyperparameters and the loss settings were selected according to the problem requirements.

\subsection{Training and testing}
As stated before, both the training and testing procedure, when possible, has been conducted under the same hypothesis and by using the same settings. All the qubits in Fig.~\ref{fig:circuit1}, Fig.~\ref{fig:bellman}, and Fig.~\ref{fig:real_amplitude} are set equal to the state $\ket{0}$.

The models were trained on the Google Colaboratory platform, where each user can count on: $1)$ a GPU Tesla K80, having 2496 CUDA cores, compute 3.7, 12G GDDR5  VRAM, $2)$ a CPU single core hyper threaded i.e (1 core, 2 threads) Xeon Processors @2.3 Ghz (No Turbo Boost), $3)$ 45MB Cache, $4)$ 12.6 GB of available RAM and $5)$ 320 GB of available disk.

Each QCNN classifier, regardless of the circuit it used, has been trained for 50 epochs, using the Adam optimizer, with a learning rate of 0.0002, and the Cross-Entropy as loss function. The two classical CNN have been trained in the same way, but they took $\sim 100$ epochs to converge.

The training procedure is summarized in Algorithm~\ref{alg:training}, where the fundamental steps of this process have been reported. The training phase, as happens for any machine learning model whose training is based on backpropagation algorithms, can be divided into two streams, the feed-forward and backward. In the first stream, input data passes through both the CNN and the Quantum Circuit, then the overall output is compared with the ground truth, to calculate the error, and through the backward stream all the model's weights are updated according to the error and its gradient. The testing of the models have been conducted on the validation dataset, according to the procedure summarized in Algorithm~\ref{alg:testing} for the sake of reproducibility.\\


\begin{algorithm}[!ht]
\SetAlgoLined
\textbf{initializeModel()}\\
\For{epoch$\leftarrow 0$ \KwTo epochs \KwBy $1$}{
    img, groundTruth  = \textbf{loadFromTrainingSet()}\\
    \Comment{Apply CNN}
    featuresMap = \textbf{applyCNN}(img)\\
    featuresVector = \textbf{flatten}(featuresMap)\\
    \Comment{Adapt features for Quantum Circuit}
    toQuantumCircuit = \textbf{applyFully1}(featuresVector)\\
    quantumOut = \textbf{applyQuantumCircuit}(toQuantumCircuit)\\
    \Comment{Adapt Quantum Output for classification}
    classification =  \textbf{softmax}(applyFullt2(applyQuantumCircuit))\\
    \Comment{Update Hybird CNN}
    error, grad = \textbf{computeErrorGrad}(classification, groundTruth)\\
    \textbf{updatesCNNWeights}(error, grad)\\
    \textbf{updatesQuantumWeights}(error, grad)\\
    }
\caption{Training of Hybrid Quantum Neural Networks}\label{alg:training}    
\end{algorithm}

\begin{algorithm}[!ht]
\SetAlgoLined
model = \textbf{loadTrainedModel()}\\
\For{img$\leftarrow$ \KwTo testing set}{
    img, groundTruth  = \textbf{loadFromTestingSet()}\\
    \textbf{append}(groundTruths, groundTruth)\\
    \Comment{Apply Trained Model}
    prediction = \textbf{applyTrainedModel}(model, img)\\
    \textbf{append}(predictions, prediction)\\
    }
\Comment{Get scores}
cm = \textbf{confusionMatrix}(groundTruths, predictions)\\
accuracy, precision, recall, f1 = \textbf{classificationReport}(groundTruths, predictions)
\caption{Testing of Hybrid Quantum Neural Networks}\label{alg:testing}    
\end{algorithm}

\section{Results} \label{sec:results}
\subsection{EuroSAT dataset classification \label{sec:eurosat-multiclass}}
In this section the results of all the proposed models are presented in the form of confusion matrices and tables with classification reports, showing accuracy, precision, recall and F1 score, as defined by equations \eqref{eqn:acc}.

\begin{subequations}
\centering
\begin{align}
    \textrm{Accuracy} &= \frac{TP+TN}{TP+FP+FN+TN}\\
    \textrm{Precision} &= \frac{TP}{TP+FP}\\
    \textrm{Recall} &= \frac{TP}{TP+FN}\\
    \textrm{F1 Score} &= 2\,\frac{\textrm{Recall}\cdot\textrm{Precision}}{\textrm{Recall}+\textrm{Precision}}
\end{align}
\label{eqn:acc}
\end{subequations}

\footnotetext{Quantum Neural Network graphics made with PlotNeuralNet~\cite{plotnet}.}
In equations \eqref{eqn:acc}, $TP, \,TN, \,FP, \,FN$ are the number of True Positive cases, True Negative cases, False Positive cases, and False Negative cases, respectively.

In Table~\ref{tab:scores1} the F1 scores are reported  for each class, together with the overall Accuracy, computed on the three proposed quantum classifier and on the two classical counterparts. While in Table~\ref{tab:scores2} and Table~\ref{tab:scores3}  the Precision and Recall are reported for each class and for each model mentioned above.

The main evident difference among the quantum-based models is the higher performance when circuits with entanglement are used, thanks to their increased computational capabilities. Both entangled circuits also performed better than the  two classical counterparts. Among circuits with entanglement, the Real Amplitudes Circuit reaches the best overall accuracy of $92\%$, a $+10\%$ gain over the second best approach. Delving into details, it has to be underlined that the model using the no entanglement circuit fails to recover the Highway class, one of the classes on which all the classifiers analyzed have found greater difficulties. This result highlights that the choice of the quantum circuit is not only linked to the type of application but also to the complexity of the data being used. In fact, this circuit has been successfully applied for digit image classification~\cite{githubquantumdl}, but its effectiveness is poor on more complex remote sensing images.

\begin{table*}[!ht]
    \centering
    \resizebox{2\columnwidth}{!}{
    \begin{tabular}{lccccccccccr}
         \toprule
         Model & & & & & F1 Score & & & & & & Accuracy\\
               & Annual Crop & Forest & Herb. Vegetation &  Highway & Industrial & Pasture & Perma. Crop & Residential & River & Sea Lake &  \\
         \midrule
         Classical v1            & 0.83 & 0.91 & 0.79 & 0.63 & 0.90 & 0.73 & 0.70 & 0.92 & 0.77 & 0.97 & 0.83\\
         Classical v2            & 0.81 & 0.94 & 0.76 & 0.67 & 0.91 & 0.83 & 0.70 & 0.89 & 0.73 & 0.95 & 0.83\\
         \midrule
         No entanglement circuit                      & 0.83 & 0.93 & 0.79 & 0.00 & 0.87 & 0.86 & 0.65 & 0.89 & 0.74 & 0.96 & 0.79 \\
         Bellman Circuit            & 0.82 & 0.89 & 0.78 & 0.72 & 0.94 & 0.78 & 0.69 & 0.94 & 0.80 & 0.97 & 0.84\\
         \textbf{Real Amplitudes Circuit} & \textbf{0.90} & \textbf{0.98} &\textbf{ 0.89} & \textbf{0.86} & \textbf{0.96} & \textbf{0.92} & \textbf{0.84} & \textbf{0.97} & \textbf{0.87} & \textbf{0.98} & \textbf{0.92}\\
         \bottomrule\\
    \end{tabular}}
    \caption{F1 Score + Accuracy}
    \label{tab:scores1}
\end{table*}

\begin{table*}[!ht]
    \centering
    \resizebox{2\columnwidth}{!}{
    \begin{tabular}{lccccccccccr}
         \toprule
         Model & & & & & Precision & & & & &\\
               & Annual Crop & Forest & Herb. Vegetation &  Highway & Industrial & Pasture & Perma. Crop & Residential & River & Sea Lake\\
         \midrule
         Classical v1            & 0.78 & 0.97 & 0.84 & 0.75 & 0.86 & 0.75 & 0.60 & \textbf{0.97} & 0.74 & \textbf{0.99}\\
         Classical v2            & 0.82 & 0.90 & 0.79 & 0.65 & 0.88 & 0.83 & 0.68 & 0.86 & 0.83 & 0.98\\
         \midrule
         No entanglement circuit      & 0.78 & 0.94 & 0.74 & 0.00 & 0.87 & 0.82 & 0.62 & 0.83 & 0.66 & 0.95\\
         Bellman Circuit            & \textbf{0.92} & 0.81 & 0.77 & 0.69 & 0.90 & 0.73 & \textbf{0.86} & 0.92 & 0.79 & \textbf{0.99}\\
         \textbf{Real Amplitudes Circuit} & 0.91 & \textbf{0.98} & \textbf{0.92} & \textbf{0.85} & \textbf{0.99} & \textbf{0.94} & 0.76 & 0.95 & \textbf{0.91} & \textbf{0.99}\\
         \bottomrule\\
    \end{tabular}}
    \caption{Precision}
    \label{tab:scores2}
\end{table*}

\begin{table*}[!ht]
    \centering
    \resizebox{2\columnwidth}{!}{
    \begin{tabular}{lccccccccccr}
         \toprule
         Model & & & & & Recall & & & & &\\
               & Annual Crop & Forest & Herb. Vegetation &  Highway & Industrial & Pasture & Perma. Crop & Residential & River & Sea Lake\\
         \midrule
         Classical v1            & 0.87 & 0.86 & 0.75 & 0.54 & 0.96 & 0.71 & 0.83 & 0.88 & 0.81 & 0.95\\
         Classical v2            & 0.80 & 0.97 & 0.72 & 0.67 & 0.94 & 0.83 & 0.71 & 0.93 & 0.72 & 0.91\\
         \midrule
         No entanglement circuit      & 0.87 & 0.92 & 0.86 & 0.00 & 0.87 & 0.89 & 0.68 & 0.98 & \textbf{0.85} & \textbf{0.98}\\
         Bellman Circuit            & 0.74 & \textbf{0.98} & 0.78 & 0.75 & \textbf{0.97} & 0.83 & 0.57 & \textbf{0.97} & 0.82 & 0.94\\
         \textbf{Real Amplitudes Circuit} & \textbf{0.89} & \textbf{0.98} & \textbf{0.87} & \textbf{0.86} & 0.94 & \textbf{0.91} & \textbf{0.93} & \textbf{0.99} & 0.83 & \textbf{0.98}\\
         \bottomrule\\
    \end{tabular}}
    \caption{Recall}
    \label{tab:scores3}
\end{table*}

\begin{figure*}[!ht]
    \centering
    \resizebox{2\columnwidth}{!}{
    \begin{tabular}{cccc}
    \includegraphics[width=0.66\columnwidth]{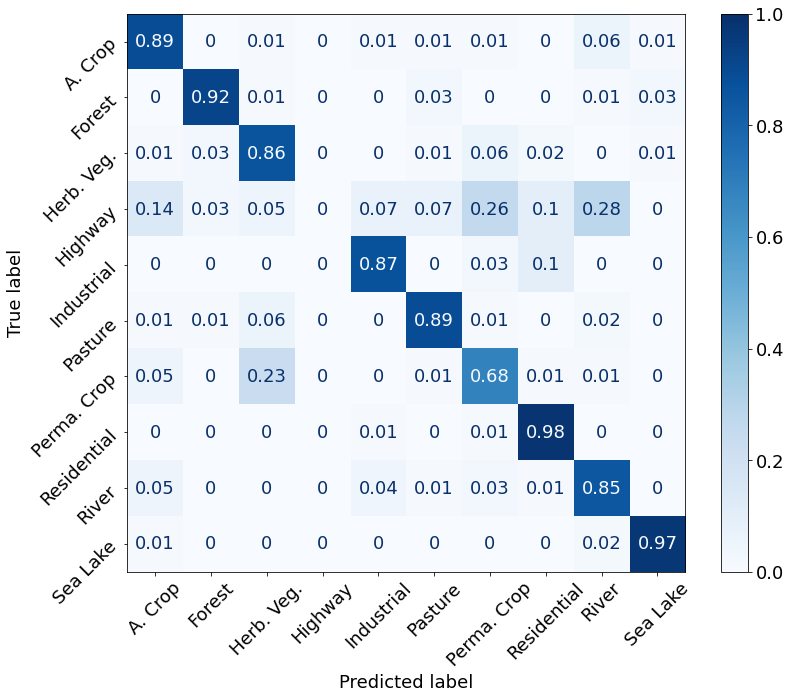} &
    \includegraphics[width=0.66\columnwidth]{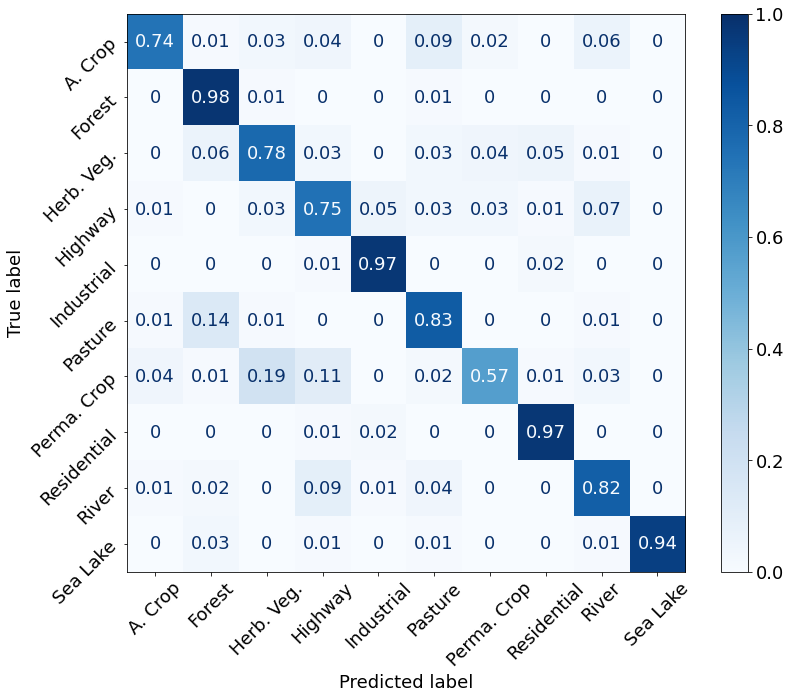} &
    \includegraphics[width=0.66\columnwidth]{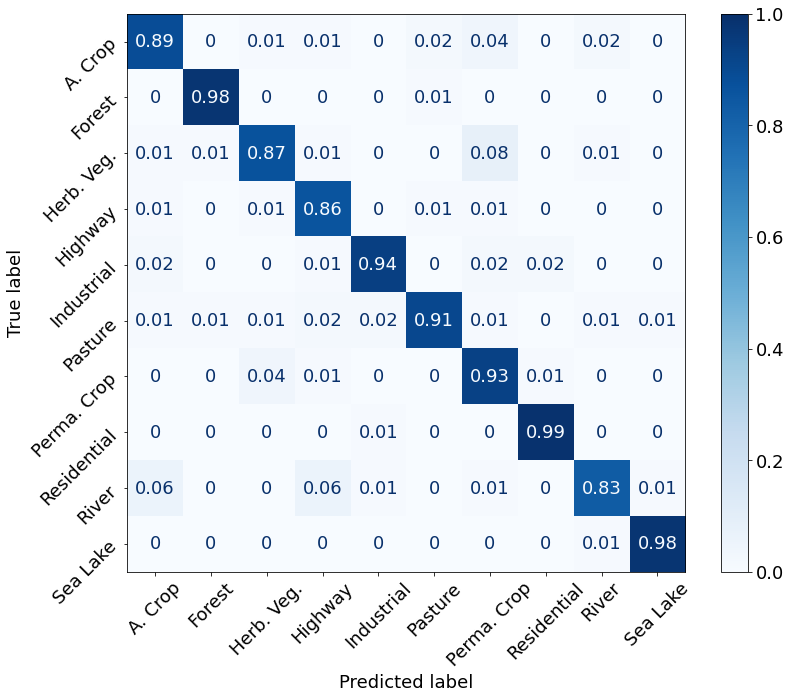} \\
    \textbf{(a)}  & \textbf{(b)} & \textbf{(c)}  \\[6pt]
    \end{tabular}}
    \resizebox{1.3\columnwidth}{!}{
    \begin{tabular}{cccc}
    \includegraphics[width=0.66\columnwidth]{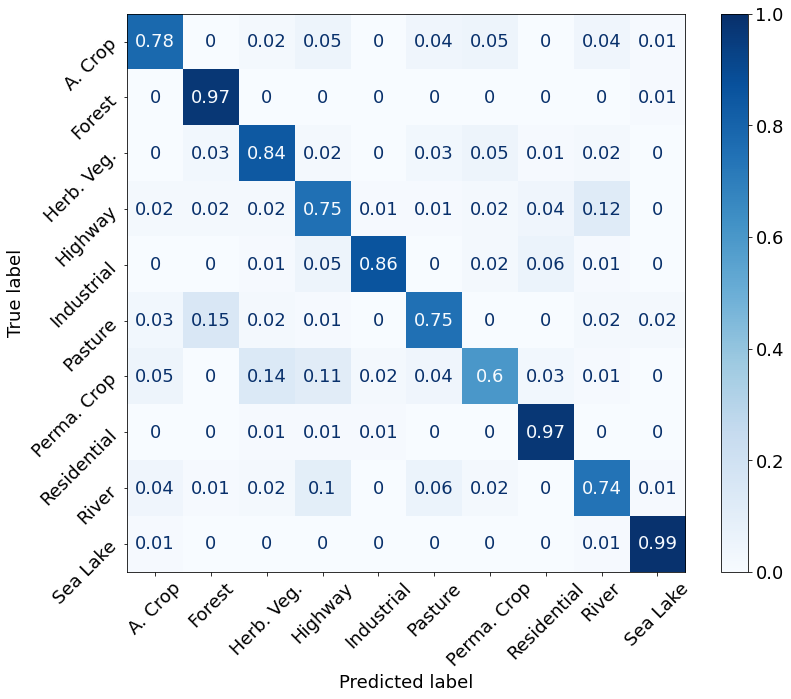} &
    \includegraphics[width=0.66\columnwidth]{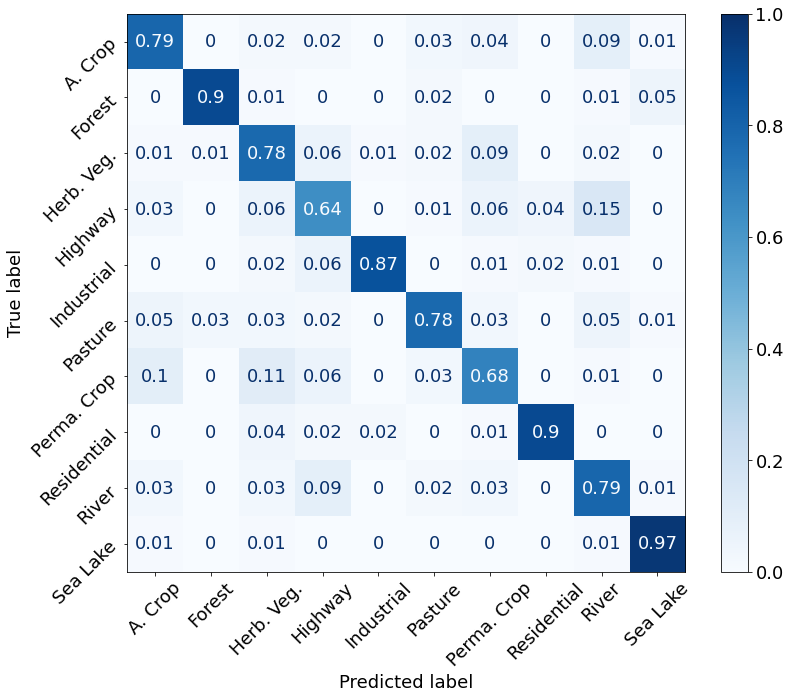} \\
    \textbf{(d)}  & \textbf{(e)}  \\[6pt]
    \end{tabular}}
    \caption{ \textbf{(a)} Confusion Matrix for no entanglement circuit
    \textbf{(b)} Confusion Matrix for Bellman Circuit
    \textbf{(c)} Confusion Matrix for Real Amplitudes circuit
    \textbf{(d)} Confusion Matrix for Classical v1
    \textbf{(e)} Confusion Matrix for Classical v2}
    \label{fig:confusion}
\end{figure*}

In Fig.~\ref{fig:confusion} the confusion matrices for each model are shown. The Real Amplitudes Circuit-based QCNN shows the best confusion matrix, with nearly-perfect scores on the diagonal. It is able to surpass the performances of all the other quantum-based models and those of the classic models, which all come up against difficulties for specific classes.

\subsection{Coarse-to-fine structured land-cover classification \label{sec:eurosat-coarse-2-fine}}

Classification results shown in section ~\ref{sec:eurosat-multiclass} and especially Table~\ref{tab:scores1} demonstrate the ability of our hybrid classical-quantum network to perform multi-class EO classification. Even if some-state-of-the-art classical networks achieve better performance (as in Helbert et al., JSTARS 2019~\cite{helber2019eurosat}),  it is worth highlighting that the proposed quantum models are extremely less complex and with very few parameters as shown in Table~\ref{tab:comparisons2}. Moreover, to further challenge the capacities of our hybrid approach of learning with a limited number of parameters, we propose a structured prediction setting, with coarse-to-fine classification, which shows on par results with the best standard approaches.

Three {\em difficult} subsets for images of visually-similar classes were created. Then, these clusters have been  used to train three hybrid QCNNs with Real Amplitudes Circuit, namely the fine-grain classifiers. In this way the 4-qubit and the entanglement have been applied within the selected macro-classes and their inherent complexity used to encode details finer than in the overall set-up. The proposed clusters are: $1)$ Vegetation:  Annual Crop, Permanent Crop, Pasture, Forest and Herbaceous Vegetation, $2)$ Urban: Highway, Industrial and Residential and $3)$ Water Bodies: River and Sea Lake.

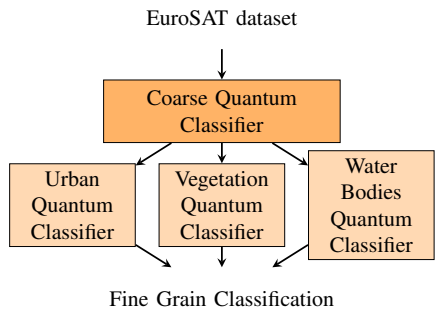
\begin{figure}[!ht]
    \centering
    \resizebox{0.65\columnwidth}{!}{
    \begin{tikzpicture}[node distance=1.5cm]
    \node (start) [startstop] {EuroSAT dataset}	;
    
    \node (coarseclass) [process,below of=start] {Coarse Quantum Classifier};
    
    \draw [arrow] (start) -- (coarseclass);
    
    \node (class2) [process2, below of = coarseclass] {Vegetation Quantum Classifier};
    
    \node (class1) [process2, left of = class2, xshift = -0.1\columnwidth] {Urban Quantum Classifier};
    
    \node (class3) [process2, right of = class2, xshift = 0.1\columnwidth] {Water Bodies Quantum Classifier};
    
    \draw [arrow] (coarseclass) -- (class2);
    \draw [arrow] (coarseclass) -- (class1);
    \draw [arrow] (coarseclass) -- (class3);
    
    \node (stop) [startstop, below of = class2] {Fine Grain Classification};
    
    \draw [arrow] (class2) -- (stop);
    \draw [arrow] (class1) -- (stop);
    \draw [arrow] (class3) -- (stop);
    
    \end{tikzpicture}}
    \caption{Coarse-to-fine land-cover classification scheme}
    \label{fig:coarse2fine}
\end{figure}

The overall structure of the coarse-to-fine land-cover classifier is shown in Fig.~\ref{fig:coarse2fine}. A first coarse classifier, based also on the real amplitudes circuit, is trained and applied to divide the data into three macro-classes. Then, based on the coarse-classifier output, the corresponding fine grain classifier is applied to obtain the final classification.

In Table~\ref{tab:coarse} the performances of the coarse classifier only are reported. The proposed model reached an overall accuracy of 98\% and an overall F1 score of 98\%.

\begin{table}[!ht]
    \centering
    \resizebox{0.75\columnwidth}{!}{\begin{tabular}{lccc}
    \toprule
    Cluster &  Precision & Recall & F1 Score\\
    \midrule
    Vegetation        & 0.97 & 0.99 & 0.98\\
    Urban             & 0.99 & 0.99 & 0.99\\
    Water Bodies      & 0.98 & 0.95 & 0.97\\
    \midrule
    Accuracy          &      &      & 0.98\\
    Macro Average     & 0.98 & 0.98 & 0.98\\
    Weighted Average  & 0.98 & 0.98 & 0.98\\
    \bottomrule\\
    \end{tabular}}
    \caption{Coarse Classification Report}
    \label{tab:coarse}
\end{table}

In Table~\ref{tab:vegetation} (resp. Table~\ref{tab:urban} and Table~\ref{tab:water} ) the performances of the fine grain classifier for the vegetation (resp. Urban and Water) classes are reported. The proposed models reached overall accuracies of $94\%$ to $99\%$ and overall F1 scores of $94\%$ to $99\%$. This is consistently better if compared with the results for each individual class obtained with the standard classifier (Table~\ref{tab:scores1}), meaning that with constant complexity on a slightly reduced dataset, the hybrid QCNN can learn finer details to distinguish similar images. 

\begin{table}[!ht]
    \centering
    \resizebox{0.75\columnwidth}{!}{\begin{tabular}{lccc}
    \toprule
    Class  &  Precision & Recall & F1 Score\\
    \midrule
    Annual Crop             & 0.93 & 0.94 & 0.93\\
    Permanent Crop          & 0.99 & 0.98 & 0.98\\
    Pasture                 & 0.92 & 0.94 & 0.93\\
    Forest                  & 0.94 & 0.89 & 0.91\\
    Herbaceous Vegetation   & 0.82 & 0.95 & 0.93\\
    \midrule
    Accuracy                &      &      & 0.94\\
    Macro Average           & 0.94 & 0.94 & 0.94\\
    Weighted Average        & 0.94 & 0.94 & 0.94\\
    \bottomrule\\
    \end{tabular}}
    \caption{Vegetation Fine Grain Classification Report}
    \label{tab:vegetation}
\end{table}

\begin{table}[!ht]
    \centering
    \resizebox{0.75\columnwidth}{!}{\begin{tabular}{lccc}
    \toprule
    Class  &  Precision & Recall & F1 Score\\
    \midrule
    Highway                 & 0.99 & 0.98 & 0.99\\
    Residential             & 0.99 & 0.99 & 0.99\\
    Industrial              & 0.99 & 0.99 & 0.99\\
    \midrule
    Accuracy                &      &      & 0.99\\
    Macro Average           & 0.99 & 0.99 & 0.99\\
    Weighted Average        & 0.99 & 0.99 & 0.99\\
    \bottomrule\\
    \end{tabular}}
    \caption{Urban Fine Grain Classification Report}
    \label{tab:urban}
\end{table}

\begin{table}[!ht]
    \centering
    \resizebox{0.75\columnwidth}{!}{\begin{tabular}{lccc}
    \toprule
    Class  &  Precision & Recall & F1 Score\\
    \midrule
    River                   & 0.97 & 0.99 & 0.99\\
    Sea Lake                & 0.99 & 0.98 & 0.99\\
    \midrule
    Accuracy                &      &      & 0.99\\
    Macro Average           & 0.99 & 0.99 & 0.99\\
    Weighted Average        & 0.99 & 0.99 & 0.99\\
    \bottomrule\\
    \end{tabular}}
    \caption{Water Bodies Fine Grain Classification Report}
    \label{tab:water}
\end{table}


In Table~\ref{tab:fine} the performances of the overall coarse-to-fine grain classifier are reported. The proposed model reached an overall accuracy of 97\% and an overall F1 score of 97\%, improving over the standard classifier by $+3\%$ and reaching performances on par with Helber et al.~\cite{helber2019eurosat} where the authors reached a 98.57\% of overall accuracy, by using a model based on the ResNet-50. It is worth to highlight that the architecture proposed in this manuscript is extremely less complex than the one proposed in~\cite{helber2019eurosat}, since the ResNet-50 is composed of 50 layers while the proposed one is composed of 6 layers only: 5 classical and 1 quantum. This is an asset for computations in environments with frugal resources.  The comparisons are better highlighted in Table \ref{tab:comparisons2}, where are reported the overall accuracy of classical and quantum models, the size of each model in terms of layers and the complexity of each model in terms of number of parameters. The table is organized in two branches, the first one containing the results of the state-of-the-art models while the second one contains the results for both the classical and quantum models proposed in this work.


\begin{table}[!ht]
    \centering
    \resizebox{0.9\columnwidth}{!}{\begin{tabular}{lccc}
    \toprule
    Class  &  Precision & Recall & F1 Score\\
    \midrule
    Annual Crop             & 0.98 & 0.93 & 0.95\\
    Permanent Crop          & 0.98 & 0.98 & 0.98\\
    Pasture                 & 0.93 & 0.94 & 0.94\\
    Forest                  & 0.95 & 0.95 & 0.95\\
    Herbaceous Vegetation   & 0.93 & 0.94 & 0.94\\
    Highway                 & 0.99 & 0.99 & 0.99\\
    Residential             & 0.99 & 0.99 & 0.99\\
    Industrial              & 0.99 & 0.99 & 0.99\\
    River                   & 0.99 & 0.99 & 0.99\\
    Sea Lake                & 0.99 & 0.99 & 0.99\\
    \midrule
    Accuracy                &      &      & 0.97\\
    Macro Average           & 0.97 & 0.97 & 0.97\\
    \bottomrule\\
    \end{tabular}}
    \caption{Coarse-to-fine land-cover quantum classifier report}
    \label{tab:fine}
\end{table}

\begin{table}[!ht]
    \centering
    \resizebox{1\columnwidth}{!}{\begin{tabular}{lcccc}
    \toprule
      Model & Overall Accuracy & N. layers & N. parameters\\
      \midrule
      Helber et Al. \cite{helber2019eurosat}  ResNet-50    & 0.98 & 50 & 25.6M\\
      Helber et Al. \cite{helber2019eurosat}  GoogleNet    & 0.98 & 27 & 7M\\
      Li et Al. \cite{li2020deep} ResNet-18                & 0.98 & 18 & 11M\\
      Sumbul et Al. \cite{sumbul2019bigearthnet} S-CNN-RGB & 0.70 & 3 & 23.584\\
      \midrule
      Classical V1                                         & 0.82 & 6 & 42.338\\
      Classical V2                                         & 0.83 & 7 & 329.290\\
      No entanglement circuit                              & 0.79 & 6 & 42.338 + 4q\\
      Bellman circuit                                      & 0.84 & 6 & 42.338 + 4q\\
      Real Amplitude circuit                               & 0.92 & 6 & 42.338 + 8q\\
      Fine land-cover classifier                           & 0.97 & 6 & 42.338 + 8q\\
      \bottomrule
    \end{tabular}}
    \caption{Comparisons with state of the art and classical methods. Table shown the model used, the overall accuracy and the number of layers to give an estimate of the complexity. All approaches in the second part of the table are our implementations, described in this article. Other comparisons with classical models can be found in~\cite{dewangkoro2021land}.}
    \label{tab:comparisons2}
\end{table}

Finally graphical results for the the Real Amplitudes Quantum Classifier and for the coarse-to-fine land cover classifier are reported in Table~\ref{tab:qualitative_realamp} and Table~\ref{tab:fine} respectively. These tables are structured in order to show correctly and wrongly predicted classes with the idea of underlying the increase of performances introduce with the coarse-to-fine structured land-cover classification.

\begin{table*}[!ht]
    \centering
    \resizebox{2\columnwidth}{!}{
    
    \begin{tabular}{l|cccccccccc}
 
    & \textbf{Annual} & \textbf{Forest} & \textbf{Herbaceous} & \textbf{Highway} & \textbf{Industrial} & \textbf{Pasture} & \textbf{Permanent} &  \textbf{Residential} & \textbf{River} & \textbf{Sea}\\
    & \textbf{Crop} &   & \textbf{Vegetation} &  &  & & \textbf{Crop} & &  & \textbf{Lake}\\
    
    &\includegraphics[width=0.16\columnwidth]{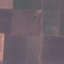} & 
    \includegraphics[width=0.16\columnwidth]{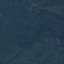} &
    \includegraphics[width=0.16\columnwidth]{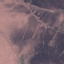} & 
    \includegraphics[width=0.16\columnwidth]{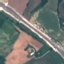} & 
    \includegraphics[width=0.16\columnwidth]{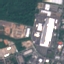} & 
    \includegraphics[width=0.16\columnwidth]{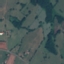} & 
    \includegraphics[width=0.16\columnwidth]{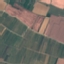} & 
    \includegraphics[width=0.16\columnwidth]{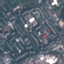} & 
    \includegraphics[width=0.16\columnwidth]{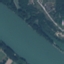} & 
    \includegraphics[width=0.16\columnwidth]{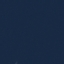}\\

    True Positive &\includegraphics[width=0.16\columnwidth]{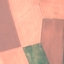} & 
    \includegraphics[width=0.16\columnwidth]{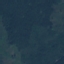} &
    \includegraphics[width=0.16\columnwidth]{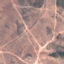} & 
    \includegraphics[width=0.16\columnwidth]{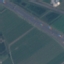} & 
    \includegraphics[width=0.16\columnwidth]{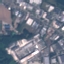} & 
    \includegraphics[width=0.16\columnwidth]{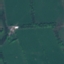} & 
    \includegraphics[width=0.16\columnwidth]{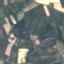} & 
    \includegraphics[width=0.16\columnwidth]{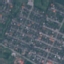} & 
    \includegraphics[width=0.16\columnwidth]{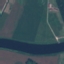} & 
    \includegraphics[width=0.16\columnwidth]{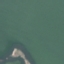}\\
    
    &\includegraphics[width=0.16\columnwidth]{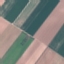} &
    \includegraphics[width=0.16\columnwidth]{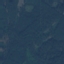} &
    \includegraphics[width=0.16\columnwidth]{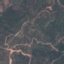} & 
    \includegraphics[width=0.16\columnwidth]{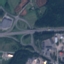} & 
    \includegraphics[width=0.16\columnwidth]{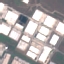} & 
    \includegraphics[width=0.16\columnwidth]{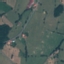} & 
    \includegraphics[width=0.16\columnwidth]{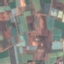} & 
    \includegraphics[width=0.16\columnwidth]{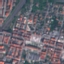} & 
    \includegraphics[width=0.16\columnwidth]{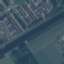} & 
    \includegraphics[width=0.16\columnwidth]{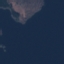}\\
    
    \midrule
    False Positive & \includegraphics[width=0.16\columnwidth]{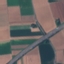}&
    ~ &
    ~ & 
    ~ & 
    \includegraphics[width=0.16\columnwidth]{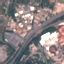} & 
    \includegraphics[width=0.16\columnwidth]{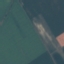} & 
    \includegraphics[width=0.16\columnwidth]{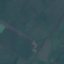} & 
    ~ & 
    ~ & 
    \includegraphics[width=0.16\columnwidth]{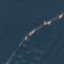}\\
    True Label & Highway &
    ~ &
    ~ & 
    ~ & 
    Highway & 
    Annual & 
    Herbaceous & 
    ~ & 
    ~ & 
    Forest\\
    ~ & ~ &
    ~ &
    ~ & 
    ~ & 
    Highway & 
    Crop & 
    Vegetation & 
    ~ & 
    ~ & 
    ~\\
 
    \end{tabular}}
    \caption{Example results of the Real Amplitudes Quantum Classifier.}
    \label{tab:qualitative_realamp}
\end{table*}

\begin{table*}[!ht]
    \centering
    \resizebox{2\columnwidth}{!}{
    
    \begin{tabular}{l|cccccccccc}
   
    & \textbf{Annual} & \textbf{Forest} & \textbf{Herbaceous} & \textbf{Highway} & \textbf{Industrial} & \textbf{Pasture} & \textbf{Permanent} &  \textbf{Residential} & \textbf{River} & \textbf{Sea}\\
    & \textbf{Crop} &   & \textbf{Vegetation} &  &  & & \textbf{Crop} & &  & \textbf{Lake}\\

    &\includegraphics[width=0.16\columnwidth]{jstars_imgs/tests/AnnualCrop_1.jpg} & 
    \includegraphics[width=0.16\columnwidth]{jstars_imgs/tests/Forest_1.jpg} &
    \includegraphics[width=0.16\columnwidth]{jstars_imgs/tests/HerbaceousVegetation_1.jpg} & 
    \includegraphics[width=0.16\columnwidth]{jstars_imgs/tests/Highway_1.jpg} & 
    \includegraphics[width=0.16\columnwidth]{jstars_imgs/tests/Industrial_1.jpg} & 
    \includegraphics[width=0.16\columnwidth]{jstars_imgs/tests/Pasture_10.jpg} & 
    \includegraphics[width=0.16\columnwidth]{jstars_imgs/tests/PermanentCrop_1.jpg} & 
    \includegraphics[width=0.16\columnwidth]{jstars_imgs/tests/Residential_3.jpg} & 
    \includegraphics[width=0.16\columnwidth]{jstars_imgs/tests/River_2.jpg} & 
    \includegraphics[width=0.16\columnwidth]{jstars_imgs/tests/SeaLake_1.jpg}\\

    True Positive &\includegraphics[width=0.16\columnwidth]{jstars_imgs/tests/AnnualCrop_4.jpg} & 
    \includegraphics[width=0.16\columnwidth]{jstars_imgs/tests/Forest_15.jpg} &
    \includegraphics[width=0.16\columnwidth]{jstars_imgs/tests/HerbaceousVegetation_7.jpg} & 
    \includegraphics[width=0.16\columnwidth]{jstars_imgs/tests/Highway_5.jpg} & 
    \includegraphics[width=0.16\columnwidth]{jstars_imgs/tests/Industrial_2.jpg} & 
    \includegraphics[width=0.16\columnwidth]{jstars_imgs/tests/Pasture_13.jpg} & 
    \includegraphics[width=0.16\columnwidth]{jstars_imgs/tests/PermanentCrop_4.jpg} & 
    \includegraphics[width=0.16\columnwidth]{jstars_imgs/tests/Residential_10.jpg} & 
    \includegraphics[width=0.16\columnwidth]{jstars_imgs/tests/River_5.jpg} & 
    \includegraphics[width=0.16\columnwidth]{jstars_imgs/tests/SeaLake_63.jpg}\\
    
    &\includegraphics[width=0.16\columnwidth]{jstars_imgs/tests/AnnualCrop_13.jpg} &
    \includegraphics[width=0.16\columnwidth]{jstars_imgs/tests/Forest_13.jpg} &
    \includegraphics[width=0.16\columnwidth]{jstars_imgs/tests/HerbaceousVegetation_13.jpg} & 
    \includegraphics[width=0.16\columnwidth]{jstars_imgs/tests/Highway_12.jpg} & 
    \includegraphics[width=0.16\columnwidth]{jstars_imgs/tests/Industrial_4.jpg} & 
    \includegraphics[width=0.16\columnwidth]{jstars_imgs/tests/Pasture_29.jpg} & 
    \includegraphics[width=0.16\columnwidth]{jstars_imgs/tests/PermanentCrop_9.jpg} & 
    \includegraphics[width=0.16\columnwidth]{jstars_imgs/tests/Residential_13.jpg} & 
    \includegraphics[width=0.16\columnwidth]{jstars_imgs/tests/River_7.jpg} & 
    \includegraphics[width=0.16\columnwidth]{jstars_imgs/tests/SeaLake_121.jpg}\\
    
    \midrule
    False Positive & ~ &
    ~ &
    ~ & 
    ~ & 
    ~ & 
    ~ & 
    \includegraphics[width=0.16\columnwidth]{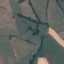} & 
    ~ & 
    ~ & 
    \includegraphics[width=0.16\columnwidth]{jstars_imgs/tests/Forest_4.jpg}\\
    True Label & ~ &
    ~ &
    ~ & 
    ~ & 
    ~ & 
    ~ & 
    Pasture & 
    ~ & 
    ~ & 
    Forest\\

    \end{tabular}}
    \caption{Example results of the coarse-to-fine quantum classifier.}
    \label{tab:qualitative_fine}
\end{table*}

~\vspace{-20pt}
\subsection{Semantic segmentation by patch-wise classification}

To further demonstrate the efficiency of the proposed approach, the trained fine-grain quantum classifier has eventually been applied to unseen Sentinel 2 images from the Onera Satellite Change Detection Dataset (OSCD)~\cite{daudt2018urban}. In order to run the classifier on these large images, we used a sliding window of 64x64 pixels, to match the size of the EuroSAT data, with a step of 32 pixels, leading to a patch-wise classification map or semantic map, reproducing the experiment of~\cite{9606737} for comparison to state-of-the-art deep learning approaches.

\begin{figure*}[!ht]
    \centering
    \resizebox{2\columnwidth}{!}{
    \begin{tabular}{c}
    \includegraphics[width=1.8\columnwidth]{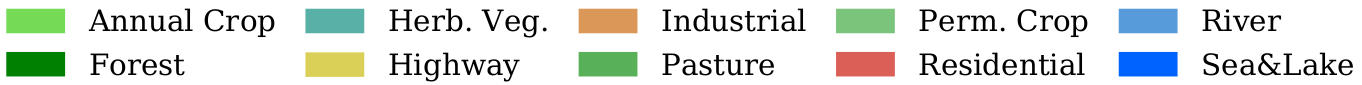}
    \end{tabular}}
    \resizebox{2\columnwidth}{!}{
    \begin{tabular}{cccc}
    \includegraphics[width=0.49\columnwidth]{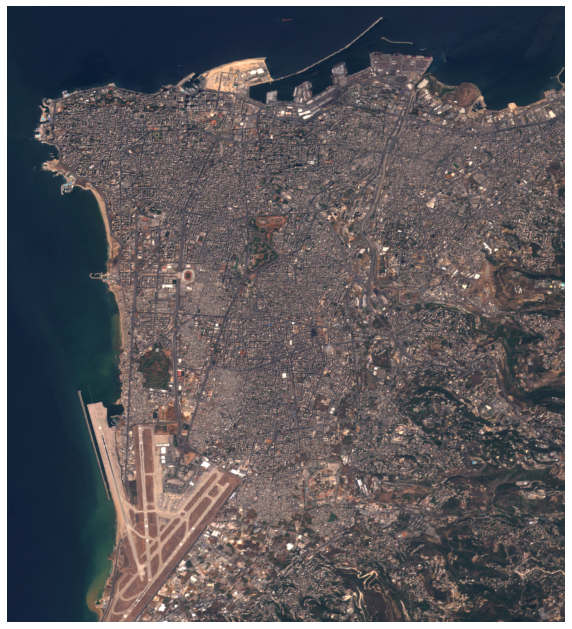} & 
    \includegraphics[width=0.49\columnwidth]{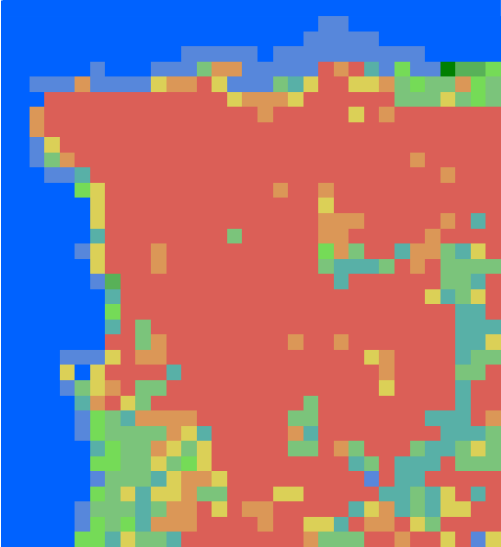}& 
    \includegraphics[width=0.489\columnwidth]{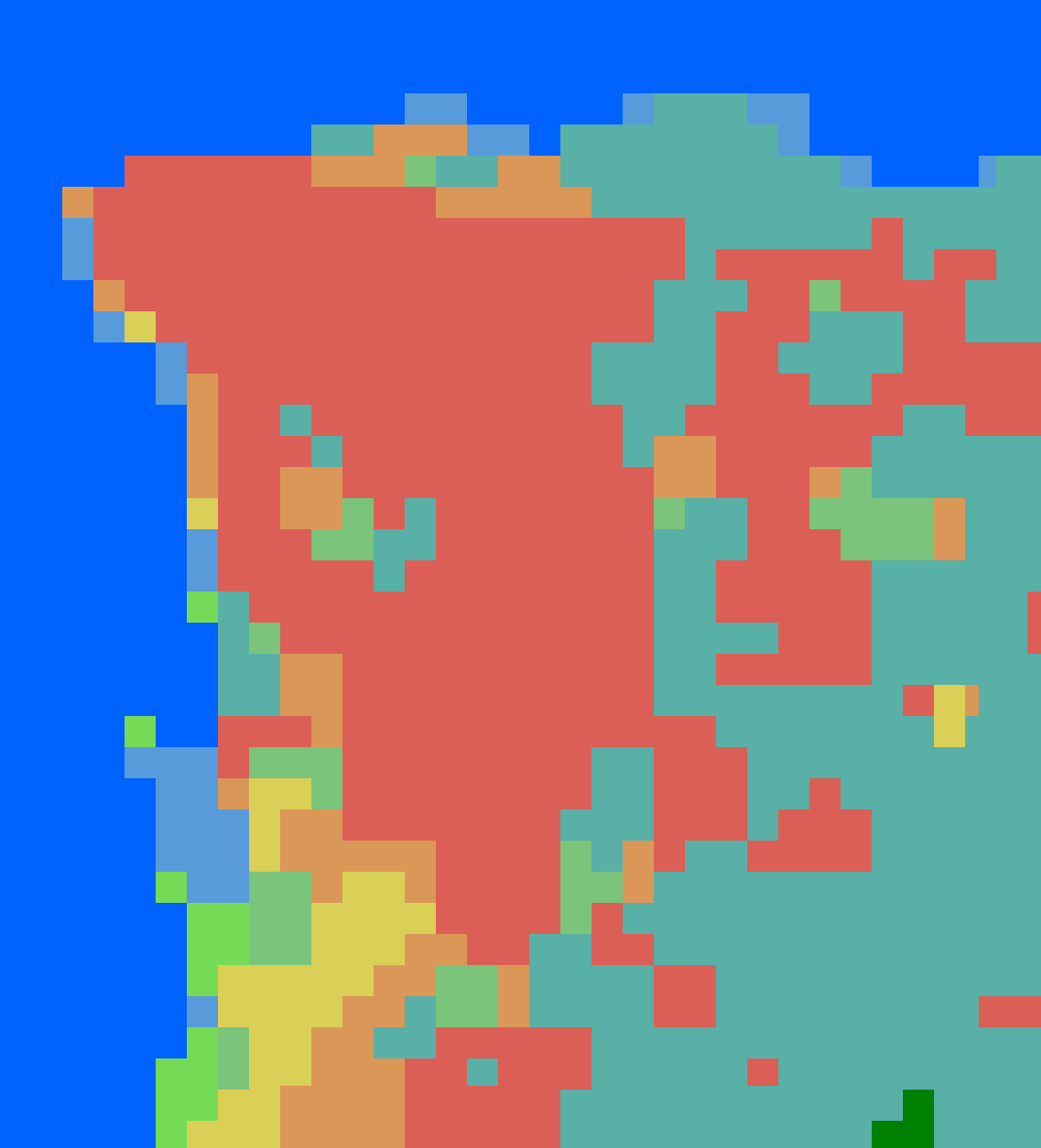} & 
    \includegraphics[width=0.489\columnwidth]{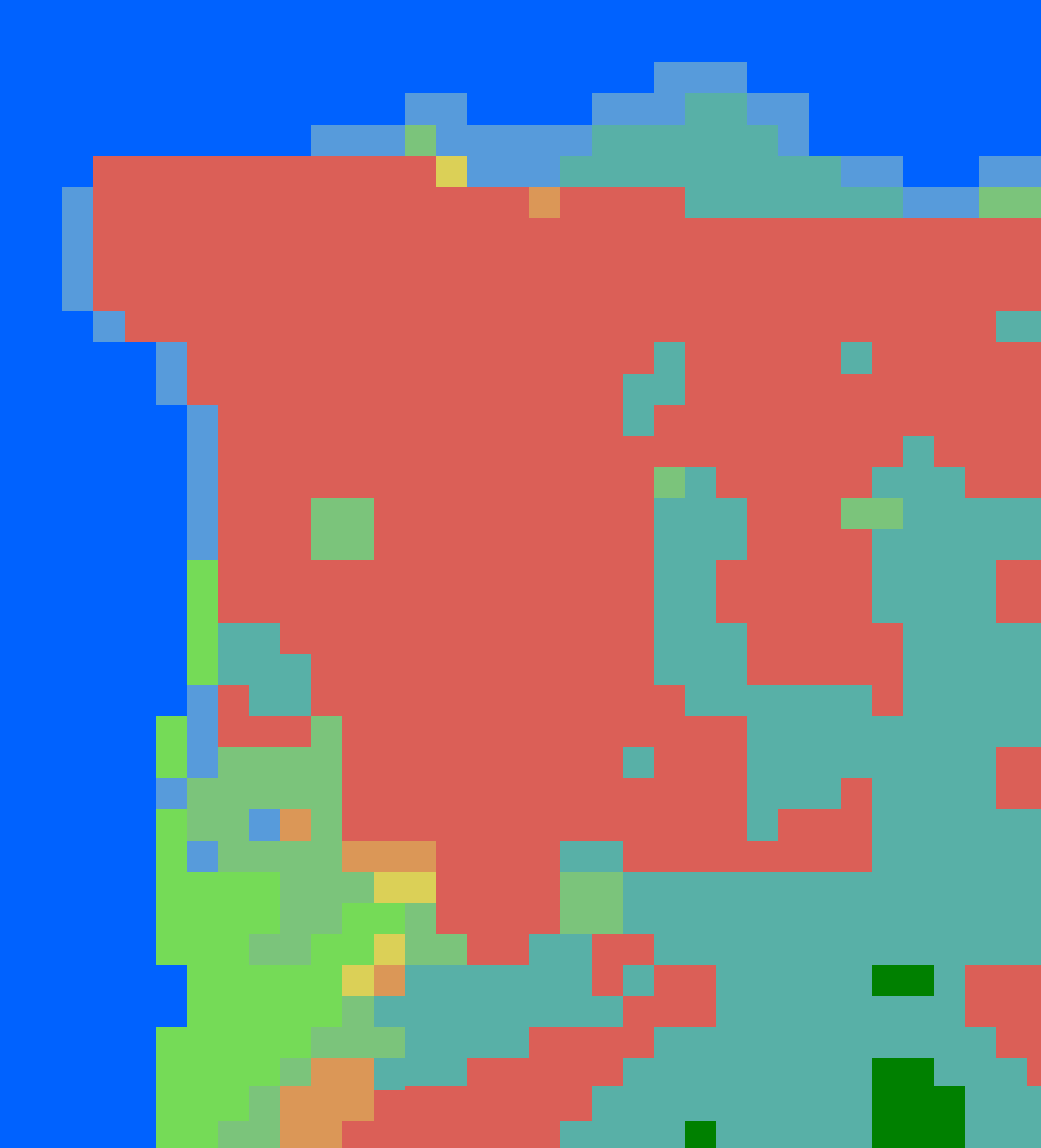}\\
    \textbf{Sentinel-2 image~\cite{daudt2018urban}}  & \textbf{coarse-to-fine quantum classifier} & \textbf{Wide-ResNet~\cite{9606737}} & \textbf{JEM~\cite{9606737}}  \\[6pt]
    \end{tabular}}
    
    \caption{LULC semantic maps on never-seen OSCD city Beirut compared with the Wide-ResNet and JEM models tested in~\cite{9606737}. \textbf{(a)} Input Image
    \textbf{(b)} Coarse-to-fine quantum classifier
    \textbf{(c)} Wide-ResNet
    \textbf{(d)} JEM}
    \label{fig:beirut}
\end{figure*}

In Figure \ref{fig:beirut} are reported the results on one location from OSCD, the city of Beirut. The maps produced by the quantum classifier have been compared with the Wide-ResNet and JEM models presented in \cite{9606737}. Results are satisfying: the classifier is able to accurately distinguish the urban, vegetation and water bodies zones along the input image. Moreover maps are comparable with other state-of-the-art solutions, with even a slight advantage on retrieving residential areas in the very urban area of Beirut.

~\vspace{-20pt}
\section{Conclusion}\label{sec:conclusion}

This paper investigates  the circuit-based hybrid QCNNs for Remote Sensing image classification.  Unlike traditional CNN architectures, the chosen QCNN updates the standard neural network with a quantum layer. The proposed method is applied to the LULC classification tasks and, through a comparative and critical analysis, the performance of different gate-based circuits has been evaluated and the hybrid QCNN has proven to be effective in terms of multiclass identification and computing efficiency.

Experiments, run on the reference benchmark EuroSAT dataset, have shown that the proposed QCNN  worked successfully   for the multiclass classification of EO scenes. Firstly, we demonstrated that the architecture with entanglement led to better results by a significant margin with respect to the others. Secondly, the quantum layer has allowed to reach better results than its classical counterpart. Moreover, all the code and experiments presented in this paper have been collected and made available open access in the GitHub page \cite{githubrepo}. This material, along with the background on QC given in this article, will hopefully be a useful tool to help the \textit{Geo-science and Remote Sensing} community tackling EO problems with this cutting-edge technology.

Regarding the classical component, which is required for data embedding given the current capacity of NISQ devices, straightforward future work will consist in exploring more powerful networks for data encoding (e.g. compressing the image information in such a way that it may be encoded on the quantum layer). Regarding the quantum component, future work will aim at increasing the proportion of quantum processing in the hybrid approach. Indeed,  more complex quantum circuits are expected to enhance the learning power of the model. In particular, quantum convolutions could be examined to incorporate spatial information and invariance in the processing. 

More fundamentally, the understanding of the probabilistic mechanisms at work in the quantum layers will represent the key to design better models, develop deep quantum learning, and eventually implement it to many real-life applications.

\section*{Acknowledgement}
Daniela A. Zaidenberg participated under a joint program of MIT and University of Sannio through the MIT Science and Technology Initiative (MISTI). This work is part of ESA $\Phi$-Lab's Quantum Computing for Earth Observation (QC4EO) initiative. We thank Pierre Philippe Mathieu Head of $\Phi$-lab explore office and Ph.D. co-supervisor of Alessandro Sebastianelli and Giuseppe Borghi Head of $\Phi$-lab, for their continual support. Moreover, the authors thank Su-yeong Chang for helpful discussions on mathematics of quantum circuits and Javiera Castillo-Navarro for sharing her expertise for the semantic segmentation experiment.


\balance
\bibliographystyle{IEEEtran.bst}
\bibliography{main}


\begin{IEEEbiography}[{\includegraphics[width=1in,height=1.15in,clip,keepaspectratio]{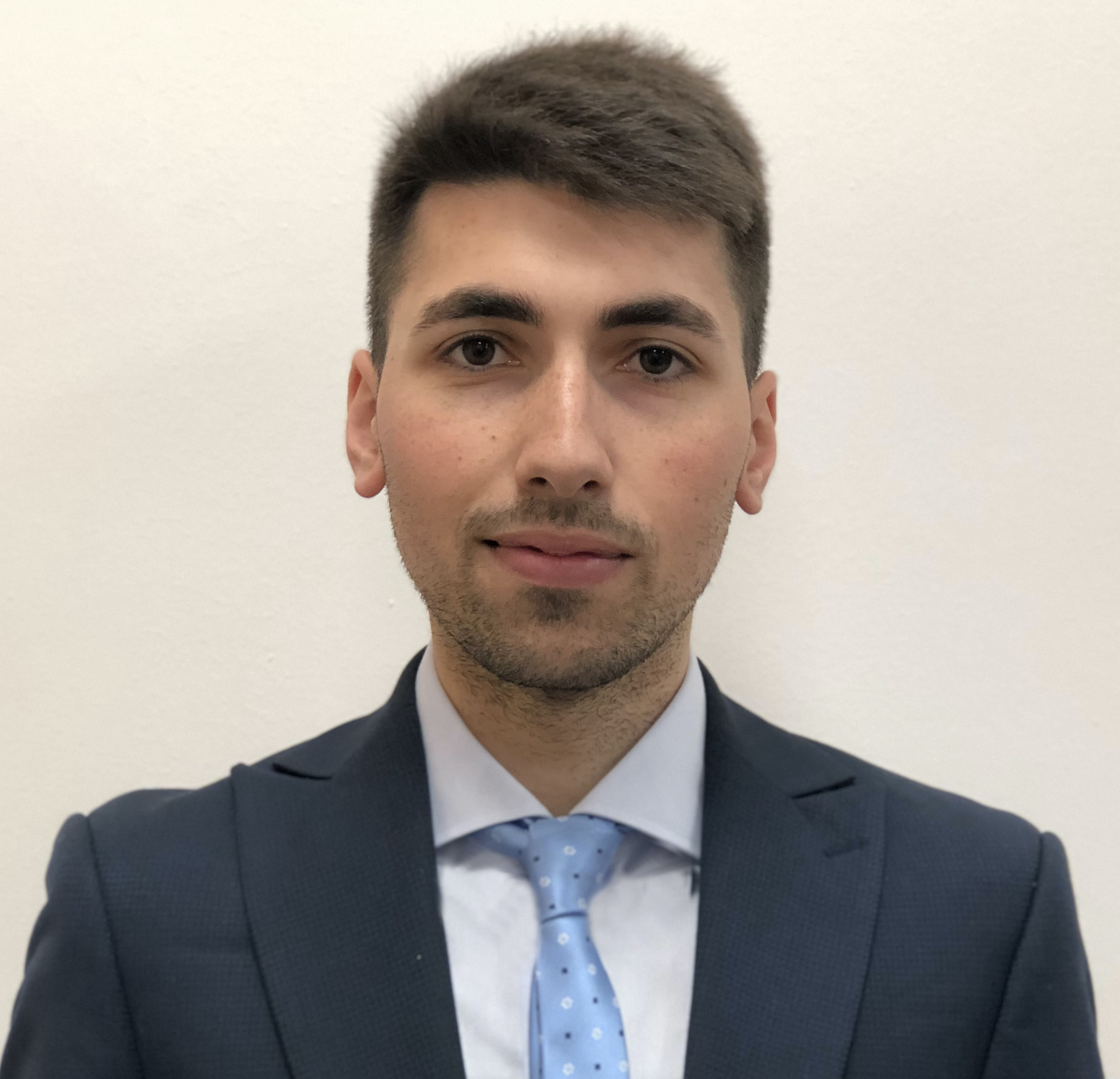}}]{Alessandro Sebastianelli} graduated  with laude in Electronic Engineering for Automation and Telecommunications at the University  of Sannio in 2019. He is enrolled in the Ph.D. program with University of Sannio, and his research topics mainly focus on Remote Sensing and Satellite data analysis, Artificial  Intelligence  techniques for Earth Observation, and data fusion. He has co-authored several papers in reputed journals and conferences for  the  sector  of  Remote Sensing. He has been a visited researcher at Phi-lab in European Space  Research  Institute  (ESRIN)  of  the  European  Space Agency (ESA), in Frascati, and still collaborates with the $\Phi$-lab  on  topics  related  to  deep  learning  applied  to  Earth Observation. He has won an ESA OSIP proposal in August 2020 presented with his Ph.D. Supervisor, Prof. Silvia L. Ullo.
\end{IEEEbiography}

\begin{IEEEbiography}[{\includegraphics[width=1in,height=1.15in,clip,keepaspectratio]{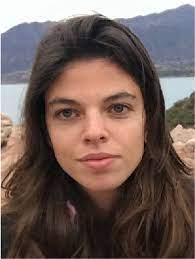}}]{Daniela Alessandra Zaidenberg} is an undergraduate researcher at MIT studying Physics and EECS with a focus on Quantum Information Science. She co-authored "Advantages and Bottlenecks of Quantum Machine Learning". She is president of the Quantum Undergraduate of MIT that serves as a pedagogical platform to teach undergrads about quantum computing, journal club, and guest speakers. Daniela is also a volunteer lecturer for qBraid a startup that teaches highschoolers and first year undergrads about quantum computing. She has worked in human computer interaction engineering, with a focus on UI design, conductive circuitry, capacitive sensing, and thermistor design. 
\end{IEEEbiography}

\begin{IEEEbiography}[{\includegraphics[width=1in,height=1.15in,clip,keepaspectratio]{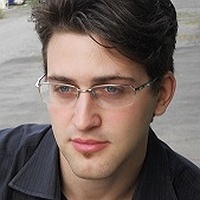}}]{Dario Spiller} is a PostDoc research fellow working for a joint research project of the Italian space agency (ASI) and the European space agency (ESA). He is an aerospace engineer with a Ph.D. in optimal control based on meta-heuristic optimization applied to space problems related to attitude and orbital maneuvers. Currently, his research is focusing on classification and regression problems applied to remote sensing test cases and solved with machine learning algorithms. He is mainly working on hyperspectral remote sensing and the PRISMA mission with application to wildfire detection and crop type classification.
\end{IEEEbiography}

\begin{IEEEbiography}[{\includegraphics[width=1in,height=1.15in,clip,keepaspectratio]{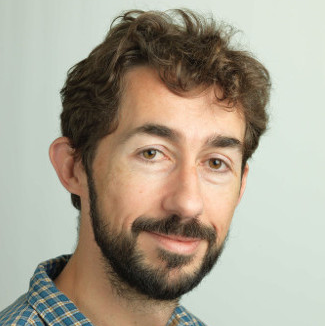}}]{Bertrand Le Saux}  (Member, IEEE) received the Ms.Eng. and M.Sc. degrees from INP, Grenoble, France, in 1999, the Ph.D. degree from the University of Versailles/Inria, Versailles, France, in 2003, and the Dr. Habil. degree from the University of Paris-Saclay, Saclay, France, in 2019. He is a Senior Scientist with the European Space Agency/European Space Research Institute $\Phi$-lab in Frascati, Italy. His research interest aims at visual understanding of the environment by data-driven techniques including Artificial Intelligence and (Quantum) Machine Learning. He is interested in tackling practical problems that arise in Earth observation, to bring solutions to current environment and population challenges. Dr. Le Saux is an Associate Editor of the Geoscience and Remote Sensing Letters. He was Co-Chair (2015–2017) and chair (2017–2019) for the IEEE GRSS Technical Committee on Image Analysis and Data Fusion.
\end{IEEEbiography}

\begin{IEEEbiography}[{\includegraphics[width=1in,height=1.15in,clip,keepaspectratio]{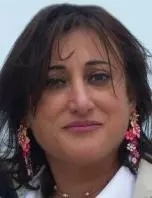}}]{Silvia Liberata Ullo} IEEE Senior Member, Industry Liaison for IEEE Joint ComSoc/VTS Italy Chapter. National Referent for FIDAPA BPW Italy Science and Technology Task Force. Researcher since 2004 in the Engineering Department of the University of Sannio, Benevento (Italy). Member of the Academic Senate and the PhD Professors’ Board. She is teaching: Signal theory and elaboration, and Telecommunication networks for Electronic Engineering, and Optical and radar remote sensing for the Ph.D. course. Authored 80+ research papers, co-authored many book chapters and served as editor of two books, and many special issues in reputed journals of her research sectors. Main interests: signal processing, remote sensing, satellite data analysis, machine learning and quantum ML, radar systems, sensor networks, and smart grids. Graduated with Laude in 1989 in Electronic Engineering,  at the Faculty of Engineering at the Federico  II  University, in  Naples, she pursued     the     M.Sc.     degree     from     the Massachusetts Institute  of Technology (MIT) Sloan  Business  School  of  Boston,  USA,  in June 1992.  She has worked in the private and public sector from 1992 to 2004, before joining the University of Sannio.
\end{IEEEbiography}

\end{document}